# Characterisation of Potential Landing Sites for the European Space Agency's Lunar Lander Project

Diego De Rosa[a]
Ben Bussey[b]
Joshua T. Cahill[b]
Tobias Lutz[c]
Ian A. Crawford[d]
Terence Hackwill[d]
Stephan van Gasselt[e]
Gerhard Neukum[e]
Lars Witte[f]
Andy McGovern[b]
Peter M. Grindrod[g]
James D. Carpenter[a]

[a]European Space Agency, Estec, Keplerlaan 1, 2200 AG Noordwijk ZH, The Netherlands
[b]The Johns Hopkins University Applied Physics Laboratory
[c]Astrium ST
[d]Birkbeck College London
[e]Freie Universität Berlin
[f]DLR Institute of Space Systems
[g]University College London

Corresponding author
Diego De Rosa
Email: diego.de.rosa@esa.int
Phone: + 31715653655
Mobile: +31654795227

## Abstract

This article describes the characterization activities of the landing sites currently envisaged for the Lunar Lander mission of the European Space Agency. These sites have been identified in the South Pole Region (-85° to -90° latitude) based on favourable illumination conditions, which make it possible to have a long-duration mission with conventional power and thermal control subsystems, capable of enduring relatively short periods of darkness (in the order of tens of hours), instead of utilising Radioisotope Heating Units. The illumination conditions are simulated at the potential landing sites based on topographic data from the Lunar Orbiter Laser Altimeter (LOLA), using three independent tools. Risk assessment of the identified sites is also being performed through independent studies. Long baseline slopes are assessed based on LOLA, while craters and boulders are detected both visually and using computer tools in Lunar Reconnaissance Orbiter Camera (LROC) images, down to a size of less than 2 m, and size-frequency distributions are generated. Shadow hazards are also assessed via LROC images. The preliminary results show that areas with quasi-continuous illumination of several months exist, but their size is small (few hundred metres); the duration of the illumination period drops quickly to less than one month outside the areas, and some areas present gaps with short illumination periods. Concerning hazard distributions, 50 m slopes are found to be shallow (few degrees) based on LOLA, whereas at the scale of the lander footprint (~5 m) they are mostly dominated by craters, expected to be mature (from geological context) and shallow (~11°). The preliminary conclusion is that the environment at the prospective landing sites is within the capabilities of the Lander design.

# 1. Introduction

The Human Space Flight and Operations directorate of the European Space Agency is conducting a mission and system study for a Lunar Lander targeting a launch date in 2018 and landing in the South Polar Region. The mission objectives are to demonstrate technologies for soft-precision landing with hazard avoidance and to conduct surface investigations in preparation for future robotic and human exploration. Carpenter et al. (2012) provide a detailed description of mission and system design, science and payload definition activities.

One of the major design drivers is the constraint not to use Radioisotope Heating Units (RHU), which are typically needed for lunar surface missions in order to survive the very low temperatures of the lunar night, which at the poles can peak to less than 100 K (Paige et al., 2009). This constraint directed the selection of the landing sites towards special locations in the lunar South Pole region (between 85° and 90° latitude south) which experience extended periods of solar illumination due to the combination of the Sun-Moon geometry and the terrain variability in the region[1].

The small inclination of the Moon's axis of rotation with respect to the ecliptic (Figure 1) causes the variations of the Sun elevation from month to month to be limited to ±1.54° over one year, as opposed to ±23.44° on Earth (Roncoli, 2005). This combines with the small elevation variations over one lunar rotation[2], equal to the colatitude of the location. This is shown in Figure 2, where the coloured lines represent the path of the Sun over one year as seen from a location at ~89.5° latitude. Blue lines correspond to the local winter, while red lines correspond to local summer. Every month, the Sun path moves from right to left, and months after month it goes upwards until it reaches a maximum (depending on the location latitude), then goes down again (dashed lines). Over one year, the excursion of the Sun elevation remains confined around 0°.

The condition described above combine with the fact that, at special locations on top of topographic heights (crater rims, ridges, mountain tops), the horizon generated by the surrounding topography can have mostly negative elevations (for example, black line in Figure 1), due to the large variations in elevation of the terrain. As a consequence, while during local winter the Sun may be completely obscured by the local topography for several days, when approaching the local spring the Sun starts being completely or even partially obscured by the highest peaks for only few days or hours. These patterns of short periods of darkness repeat each month, with decreasing duration, until the Sun becomes fully visible for several months.

The availability of solar power at these locations allows a surface mission duration of more than 14 days and potentially several months, provided that the area over which these illumination conditions occur can contain the landing dispersion ellipse, i.e. the area on the lunar surface where the lander can touchdown with a given probability (usually 99.7%). In addition, if the lander's power and thermal control subsystems can endure the short periods of darkness, the surface mission phase can be further extended.

---

[1] A dedicated analysis (not reported here) showed that the illumination conditions in the north polar region are not as good as those offered by the south polar region, due to lower terrain variations in the northern region.
[2] More precisely one synodic period, i.e. the time interval between two consecutive passes of the Sun at the same local azimuth, equal to ~29.53 days.

Another mission constraint is the unavailability of an orbiter for communications relay. As a consequence, the potential landing site must have sufficient visibility of Earth to be able to send engineering and scientific data and to receive commands. This is generally achieved at the potential landing sites at the poles, where Earth visibility has a windows of approximately 14 days, due to the fact that the Earth centre follows a pattern of approximately ±6.5° in elevation every months, constrained to approximately ±8° in azimuth. The ~14 day windows however demands a relatively high level of autonomy for the on-board surface data handling system.

In addition, surface hazards can exist at the potential landing sites, which can cause the lander to capsize or can damage the lander structure at touchdown. In order to cope with the possible presence of hazards, the lander carries on-board an autonomous Hazard Detection and Avoidance system, capable of identifying surface hazards and performing a retargeting manoeuvre towards a safe point, if necessary. The distribution of hazards within the landing site must be compatible with the capability of the HDA system and of the lander.

In order to assess the feasibility of the scenario described above and to evaluate the impacts on the mission and system design of the environment at the provisional polar landing sites, a detailed characterization of the landing site has been started. The characterization process consists in analysing the available lunar data sets, mainly from the instruments on-board the Lunar Reconnaissance Orbiter (LRO); developing and validating the necessary tools for analysis and simulation; generating all data products that are needed to characterise the landing site environment in terms of the illumination conditions and the distribution of terrain features that can endanger the landing and surface operations. This article presents the status of the analyses, performed to date, on the illumination conditions and hazard distributions at the landing sites. Further steps are needed for a complete certification of the landing sites and are outlined in Section 4.

## 2. Characterization of the illumination conditions

Several studies (Bussey et al., 1999, Fincannon, 2007, Noda et al., 2008, Bussey et al., 2010, Mazarico et al., 2011a) have been performed in the past in order to identify the polar locations that offer the most favourable illumination conditions for solar power generation and benign temperature environment, and to quantify the amount of illumination that they receive. However, the level of detail at which these studies were performed did not allow a precise estimate of the physical characteristics of these areas regarded as landing sites, in particular in terms of extent of the areas and uniformity of the distribution of illumination within each area. In addition, the parameters used as a measure of the illumination conditions did not pertain to the design case of the Lunar Lander, being for example expressed as total accumulated time of Sun visibility over one year or longest period of darkness.

This motivated, in the frame of the Lunar Lander project, the work of Vanoutryve et al. (2010) to characterise for the first time the illumination conditions at specific sites, in sufficient detail and using parameters directly relevant to system and mission design. This study was based on altimetry data from the Laser Altimeter aboard the SELENE spacecraft (Araki et al., 2009). In a continued effort and in the frame of the Lunar Lander project, the illumination conditions of the potential landing sites are being characterised through computer simulations based on topographic data from the Lunar Orbiter Laser Altimeter (LOLA), using independent tools at Astrium Space Transportation (Bremen), The John Hopkins Applied Physics Laboratory and ESA.

The objective of this analysis is to confirm the location of potential landing sites offering favourable illumination conditions and to derive the characteristics of these sites related to illumination, which have an impact on the lander's design. The primary result is an estimate of the size of the areas with favourable illumination conditions, which drives the requirement of the landing dispersion of the system. In parallel, the duration of surface mission phase can be estimated from the duration of the periods of illumination interrupted only by short periods of darkness. The required solar array height to obtain these conditions is also derived, as well as the required performances of the power and thermal control subsystems, in terms of the duration of the darkness periods that must be sustained. The analyse shall also derive the Earth communication windows and combine them with the illumination periods, as an input to mission planning and definition of the surface operations.

In order to be able to assess the probability of mission success, the characteristics described above need to be derived with a given confidence level, which is related to the general requirement of probability of mission success and to the concept of landing site certification. This is achieved by taking into account the errors of the terrain models used for simulation and by validating the computer tools.

## 2.1  Illumination modelling

The tools used for the analyses simulate the illumination conditions by first computing the local horizon for a set of points within a Region of Interest (RoI). The horizon is computed in general using a Digital Elevation Model (DEM) of the surrounding topography, but the method with which the DEM is used changes slightly from tool to tool (see below). The tools can simulate different horizons taking as an input the height above the surface, which corresponds to the height of the solar arrays, or more precisely to the lower end of the solar array strings (the solar arrays are mounted on the lateral body of the lander, see Carpenter et al, 2012, and the strings run vertically, dividing the panels in two horizontal parts).

The Sun position is then simulated using ephemeris data (JPL HORIZONS, 2012), over a simulation period of 12 synodic periods (~29.53 days), starting from the lunar southern winter solstice. This is defined either as the epoch of maximum sub-solar latitude or as the epoch halfway through a northbound and southbound equator crossing of the sub-solar point, according to different conventions. Table 1 reports the dates of the lunar southern winter solstice for the period of interest for this study. Start date and end date of the simulation periods differ by up to 15 days in the considered years, depending on the definition, but the impact on the results is negligible.

The illumination is then simulated, either in terms of visibility of the Sun centre point or in terms of the fraction of the visible Sun, using different methods. Figure 3.a shows the time history of the visible Sun fraction, for one simulation period, computed every hour. The repeating patterns of darkness and illumination are visible. The visible Sun fraction time history at a given site is converted to a binary illumination/darkness pattern (Figure 3.b) by applying a threshold, which is roughly proportional to the power needed to operate the surface payload. In the example, 50% of Sun fraction or more corresponds to the binary illumination state. Short illumination periods are pre-filtered (Figure 3.c) by joining the preceding and following darkness periods into a single darkness period. This is done in order to exclude unfeasibly short battery re-charge times. Short periods of darkness are also filtered out (Figure 3.c); they correspond to the time that can be sustained by the power/thermal control sub-systems.

The final result of this thresholding and filtering operations is the duration of the Longest Quasi-Continuous Illumination Period (LQCIP) for the point being analysed, i.e. the longest illumination period in Figure 3.d.

The procedure described above is repeated for all points within a Region of Interest. The duration of the LQCIP is mapped, typically using a polar stereographic projection (Snyder, 1987). Figure 4 shows an example LQCIP map for the Connecting Ridge area (see Figure 5). The LQCIP map reports the LQCIP duration, color-coded in days, for 2 m height above the surface, a filter for short darkness periods of 60 hours, a filter for short illumination periods of 10 hours, and year 2019. The spacing of the points within the RoI (referred to also as the analysis resolution or gridding and corresponding to the pixel size of the map) is selected based on the desired level of granularity of the analysis and considering the limitations of the available data, and is 40 m in the example (see Section 2.2).

Groups of pixels in the LQCIP maps which have a long LQCIP duration form a potential landing area. The LQCIP duration within this area gives an indication of the potential duration of the surface mission phase, with the caveat that LQCIP's may not overlap completely for points in one landing area, and this may lead to slightly optimistic conclusions on the surface mission phase duration for one landing area. In order to quantify this effect, we performed the following check: we took the pixels with LQCIP duration longer than a minimum required (corresponding to the points with acceptable surface mission duration, with the assumption that these are the pixels forming the landing area); among these pixels, we found the pixel with the latest start of the LQCIP ($t_{ls}$) and the pixel with the earliest end of the LQCIP ($t_{ee}$); we computed the time elapsed from $t_{ls}$ to $t_{ee}$ and verified that this time interval does not differ by more than 3% of the minimum required LQCIP. This implies that, if we plan for a landing date not earlier than $t_{ls}$, the actual surface mission duration will be 3% less than the minimum required LQCIP, in a worst case in which we land in the pixel with LQCIP end at time $t_{ee}$.

Several LQCIP maps can be produced, varying mainly the duration of the darkness periods and the height above the surface at which the illumination is computed, and used for design purposes. Conditions of direct communications to Earth are simulated in a similar manner, using the Earth centre or a ground station as sources of visibility conditions. The determination of the communication windows is also crucial to determine the actual duration of the surface mission phase, the possible start of the surface operations and therefore the opportunities for landing.

In this study, we focused on well determined Regions of Interest, known from literature and past studies to have good illumination conditions (Vanoutryve et al. 2010). Figure 5 shows a map of the South Polar Region based on a LOLA DEM with the location of these RoI's. Table 2 gives the coordinates of the centre of the RoI (in the Mean Earth/Polar Axis reference system, Roncoli, 2005) in geodetic coordinates and polar stereographic projection. Three independent tools are being used for the analyses. They are described in the following.

### 2.1.1 Astrium ST Illumination and Communication Analysis Tool

The Illumination and Communication Analysis Tool (ICAT) is a complete new development in MATLAB® (MATLAB, 2010) and C language, which was started for the Lunar Lander project during phase A. ICAT generates a horizon profile for each point within a RoI with a "brute-force" method, i.e. by sampling elevation data from the DEM. The samples are taken along 720 azimuth lines (0.5 degree spacing), i.e. lines lying on the plane normal to the local vertical and starting at the current observer location. A 0.5° spacing between azimuth lines has been selected due to the fact that the Sun takes approximately 1 hour (the simulation step) to cover its own size, which in average is slightly more than 0.5°. The sample are taken at variable ranges along the azimuth lines, i.e. points within 3 km are sampled every 20 m, points further than 75 km are sampled every 2 km and for intermediate ranges a linear variation of the sampling distance is used. This approach takes into account the possible altitude variations of the terrain over a given baseline, and the fact the angular size of these variations decreases with the distance from the observer location. As a consequence, less samples are needed further away from the observer in order to obtain the same vertical accuracy, with a gain in computing time (being the data access operation the most consuming process in the horizon computation). The maximum range at which samples are taken is 150 km, which has been estimated as the range at which the highest terrain point in the region has an elevation lower than the minimum Sun elevation (e.g. -6.5° at -85° latitude). Once the values of the local horizon are established, they are stored for later assessment of different time periods, as the most CPU time consuming operation is the establishment of the horizons.

The second step is the establishment of the illumination or communication patterns. The Sun and Earth ground station positions are stored in ephemeris files derived from JPL's Horizon database (JPL HORIZONS, 2012). Data was derived for Moon centred coordinates and is corrected to the current observer position to account for parallax effects. The patterns are established by evaluating whether the Sun (or the ground station in the case of the communication analyses) is above or below the local horizon at every time step. The exact elevation of the horizon at the Sun azimuth given by the current simulation time is obtained by linear interpolation of the elevation values at the surrounding points of the horizon (Figure 6). The Sun fraction above the horizon is evaluated by assuming a Sun disc with an angular diameter of 0.533°. Only the vertical fraction of the visible Sun is computed (see Figure 6).

Once computed, the illumination patterns are stored for later analyses. As the "brute-force" approach of ICAT leads to a significant computational effort, the tool is built to run in a distributed computer architecture. A master computer takes the user inputs and separates the computation into smaller work packages. The packages are distributed to the computers in the network and processed. The results are reported back to the master where they are merged into consolidated output files.

### 2.1.2 ESA's Coverage tool

ESA's Coverage tool is based on AGI Satellite Tool Kit (STK; Satellite Tool Kit, 2010) and MATLAB. MATLAB is used for pre-processing of the terrain data, to receive the simulation parameters from the user and to control STK, using the connect scripting language provided by the STK Integration Module. STK ingests the terrain data, computes the horizon for each location requested by MATLAB and computes an access, i.e. the presence of a direct line connecting a landing site and the Sun (for illumination) or Earth (for communication) which is unbroken by the lunar terrain, for each time step in the requested period. MATLAB is then used to store, post-process and display the results.

STK has been selected because it is a widely used tool with a proven record of usage in different applications (Satellite Tool Kit, Case studies, 2012). However, it had to be tuned in order to account for the higher curvature of the lunar surface compared to the Earth, for which STK is optimised, and to obtain an horizon with a higher resolution, compared to the default settings (1°). The biggest limitation of STK is that the Sun is simulated only as a point source. However, this is roughly equivalent to having a 50% threshold for the illumination/darkness transition, and the impact on the results is secondary compared to the other parameters, in most cases (see Section 2.3).

### 2.1.3 APL LunarShader

LunarShader is a computer program implemented in Java and C that was developed at APL. The tool can be run in a 'forward looking' mode, where rays progress from the Sun across the terrain paths or it can start with a specific observer point and work in a 'back looking' mode towards the Sun. The analyses performed for this study are based on the 'forward looking' mode.

In 'forward-looking' mode, the program implements a ray tracing approach to finding shadows on terrain surfaces. The basis of the ray tracing algorithm is simple vector geometry: we construct a vector from the centre of the illumination source (centre ray) to a point on the terrain and then project the vector past the point and find the next terrain point it intercepts. The first point of the terrain is lit and is established as an horizon point. If the vector from the horizon point to the next point is above the centre ray, the new point is lit and is set as the new horizon point. If the vector from the horizon to the new point is below the centre ray, the new point is either in shadow or partially illuminated, and it is not an horizon point. The fraction of Sun that is visible from the new point is computed by measuring the angle between the vector from the horizon to the new point and the top ray (the vector from the top of the illumination source) and by multiplying this by a sigmoid function, to take into account the Sun disc. The process is repeated until the vector no longer intercepts the surface. Figure 7 shows the process of running the ray trace algorithm along a single terrain path. The process is performed on equally spaced points along the ray, and the elevation of the terrain is derived at each point on the ray from interpolation of the four terrain points surrounding the point on the ray. In three dimensions, rays are traced parallel to each other (for a given Sun azimuth) such that the spacing between rays is slightly less than half a pixel.

The tool in back-looking mode follows a similar approach to trace a ray from a specified location to the Sun and find the horizon as seen from that location. It is slower because it deals with one point at a time but it is slightly more accurate, particularly for partial Sun visibility, and because it returns the result for the exact pixel location of the point being analysed. This mode is used for determining the height at which the Sun is visible from a given location, which is recorded into a spreadsheet. From this spreadsheet, the durations of eclipse and sunlit periods for a point at a given height can also be derived. This mode is also better suited for calculations of individual points.

## 2.2 LOLA data analysis

All the simulations run for this study make use of Digital Elevation Models derived from the LOLA instrument. LOLA is a five beam laser with a spot patter and a repetition frequency such that measurement 25 m apart from each other can be collected along each individual track (Smith et al., 2010). As a consequence of LRO polar orbit, the average distance between LOLA tracks in the east-west direction is in the order of 1-2 km at the equator and decreases at higher latitudes. Measurements are generally very dense at polar sites (down to few metres spacing between measurements) but also non-uniform, presenting gaps of several tens of metres at some locations. This is visible in Figure 8, where a hillshade of the LOLA DEM for the Connecting ridge area is superimposed to the individual LOLA measurements, colourised by track number. An example

LQCIP map is also given for reference. Measurements for sites around -85° latitude are less dense, as shown in Figure 9, where a lateral spacing of several hundreds of metres can be observed.

LOLA measurements have a precision between (8.4, 8.2, 1.6) m and (10.67 10.09 1.13) m in the Along Track, Cross Track and Radial directions respectively, estimated as the root-mean-square deviation of the LRO orbit solutions, after cross-over analysis and depending on the orbit solution (Mazarico et al., 2011b). The absolute accuracy of the orbit solutions, and therefore the error of the true position of the terrain point, is estimated to be of the same order of magnitude.

LOLA Digital Elevation Models (LDEM products), which are used for this analysis, are built by binning all valid measurements into the map grid cells and taking the median of measurements within one cell, if more than one exist (Smith et al., 2010). After binning, the terrain is interpolated on a regular grid, using the "surface" function of the Generic Mapping Tools package (Wessel and Smith, 1991), as described in the LDEM label files. Table 3 shows data density within the selected Regions of Interest, in terms of percentage of LDEM pixels having at least one measurement, for different LDEM resolutions and coverage, as obtained by the data count files (Neumann et al., 2011) of LOLA release 8.

As a consequence of combined spot density and absolute accuracy, we don't use LDEM products with resolution higher than 40 m. In addition, we assume that locally we can use with confidence LDEM for which at least 75% of the pixels have one measurement, depending on the location being analysed. These assumptions are supported, at least locally, by the fact that small features (e.g. craters of 60 m in diameter) can be clearly resolved in the data (see discussion of Figure 13). It is understood that local effects of nearby terrain features at resolutions smaller than 40 m will not be captured. The effect of missing measurement at points where the data is interpolated will be modelled as errors (see Section 2.2.1).

LOLA maps also present evident artefacts, highlighted by the hillshade on the top of Figure 8. These artefacts are measurement tracks with considerably higher position errors. We verify a-posteriori that these artefacts do not generate false results in the simulations.

### 2.2.1 Error modelling

The final objective of the illumination analyses is to derive the characteristics of the landing site related to illumination (size of the areas, LQCIP duration) with a required confidence level. The quantification of this confidence level shall identify the contribution of the illumination conditions to the general mission success probability. As a consequence, part of this process is to take into account the known sources of errors in the simulation due to the input data. In the following, we outline possible error models.

A simple error model consists in adding 3 times the standard deviation to the altitude value of each terrain map pixel surrounding the pixel being evaluated and calculate the illumination conditions. This would provide illumination conditions with a confidence level of 99.7%. However, this model returns unfeasible illumination conditions for the lander, when comparing the size of the lander, and in particular the solar arrays height (~2 m), to the current data errors identified in Section 2.2. In addition, this model is over-conservative, in that it doesn't take into account the autocorrelation of the terrain.

Another (ideal) solution for the error model would be to run the illumination simulations in a Monte-Carlo fashion, by applying a random dispersion to the terrain point altitude at each run, with the given standard deviation, repeating the simulation N times, taking for each result (for example LQCIP duration) the 3 sigma deviation around the average of the simulation set, in a worst sense (e.g. the shortest LQCIP). This solution however is not practical since it involves very large computation resources.

In addition, each terrain LDEM point error may be modelled differently, depending on the number of LOLA samples within one grid element of the DEM and on the quality of Orbit Determination solution associated to single LOLA measurements.

Work on error modelling is currently on-going but is not presented here. It will be presented in the future, also in view of the possible improvements of the LDEM products.

## 2.3 Preliminary results

Illumination conditions at the RoI's identified in Section 2.1 have been simulated using the Astrium ICAT and ESA Coverage Analysis tools. The simulations were run for year 2019, corresponding to a launch date in 2018 for the Lunar Lander mission[3]. The resolution of the input terrain maps and consequently the spacing of the analysed points within each RoI were selected based on the analysis of Section 2.2, which translates into a map resolution of 40 m for the sites within 2° of latitude from the pole and a resolution of 80 m for remainder of the sites. We used DEM's covering up to -75° latitude, which was found to be the maximum theoretical distance at which the highest point in the region can mask the Sun, as seen from sites around -85° latitude.

The preliminary results of the illumination analyses show that the most promising areas with LQCIP duration of several months are on the ridge connecting the Shackleton and de Garlache craters (Connecting Ridge, CR1), on the Leibnitz-β plateau (LP1), on the de Garlache rim (GR1), on the Malapert massif peaks (MP1 and MP2) and on the Shackleton crater rim (SR1). They are shown in Figure 10.

The results also show that the size of the areas with long LQCIP duration is in the order of few hundreds of metres. This can be expected, considering that a basic requirement for an area with good illumination at the poles is to be relatively flat but also limited in extent, in order for the area not to shadow itself when the Sun elevation is small (near zero degrees). In addition, some of the areas are located on the top of very narrow topographic height, such as the rim of the Shackleton crater. The areas present elongated shapes, which could accommodate larger dispersions along-track, at the cost of constraints at mission level in terms of direction of the approach path and consequently descent orbit (Carpenter et al., 2012). LQCIP duration ranges from 4 to 9 months within the area (using typical design parameters of 60 hours and 2 m height), but drops quickly to less than one month outside. This can also be expected, as the topographic feature experiencing favourable illumination conditions (hill, rim etc) shadows itself.

An extensive sensitivity analysis to the simulation parameters has been performed. The LQCIP maps in Figure 11 show the sensitivity of the extent of the areas with favourable illumination and the sensitivity of the LQCIP duration to the height above the surface and to the darkness periods duration. The analysis shows that increasing the height above the surface, for example by mounting the solar arrays on a deployable mast, improves greatly both the size of the areas and the mean duration of the LQCIP within the area. The size of the area is sensitive to a lesser extent to the duration of the short periods of darkness, which has a more significant impact on the mean LQCIP duration. The threshold used to convert the visible Sun fraction time history to a binary illumination/darkness pattern doesn't have a major impact on the duration of the LQCIP (~3% of LQCIP) and the size of the areas (see Figure 12). Short illumination periods between long darkness periods are not frequent and don't impact significantly the size and the LQCIP duration (their effect on the LQCIP maps is not visible).

---

[3] Results change slightly with the year (Vanoutryve et al. 2010). This is due to the fact that the rotation of the Moon and the Earth revolution are asynchronous and, due to the slow rotation of the Moon, the elevation of the Sun for a given azimuth and a given day after solstice can change significantly. As a consequence, the simulation should be repeated especially to account for variations in the exact dates of start of the LQCIP.

It was also found that some areas present gaps with short LQCIP durations. This fact is being addressed by comparison of the LQCIP maps with images of the sites taken by the Lunar Reconnaissance Orbiter Camera (LROC) (Robinson et al., 2010), mainly by the Narrow Angle Camera (NAC). By comparing LQCIP maps with real images, we check visually whether the gaps are artefacts in the terrain data or whether they are real terrain features. In this case, we want to check whether the gaps are permanently shadowed or whether they can be potentially illuminated at time of landing. In the latter case, the lander may land inside these gaps and experience long periods of darkness. Figure 13 and Figure 14 show a preliminary result of this analysis: a cut-out of two NAC images of the RoI with complementary illumination angles (M112490681RE and M118429312LE), which allow a visual assessment of almost all the terrain features in the region. The preliminary conclusion is that the gaps indeed match with real terrain features. These are mainly degraded craters (60 to 120 m in diameter) that can be expected to be always in shadow, when considering: the low Sun elevation; a degraded crater model (11° slope from bottom to top of the rim, Melosh, 1989); the slope of the underlying terrain (few degrees, see Figure 21). This is also supported by the fact that these gaps disappear when increasing the height above the surface to e.g. 5 m (Figure 11)

The results of the simulation of direct to Earth communication show that communication windows generally follow a regular pattern of ~14 days. As expected, areas around -85° latitude show a slightly longer window, up to ~19 days (LP1). For none of the RoI it was found that elevated terrain features exist towards the Earth direction, which may cause a significant decrease in Earth visibility. The effect of local terrain were also taken into account by imposing a 2° minimum elevation constraint. The results show a reduction of maximum 5% of communication availability.

## 2.4 Validation

The analysis method described above is being validated through a combination of cross-validation of the tools, comparison of the output of the simulations with results from literature and comparison of the output with LROC images of the surface (mainly NAC images). The objectives of the validation is to verify that the results of the different tools agree, to find the limits of the approximations in each tool and to assess the limitations of the input terrain data.

We compared the results from ESA Coverage Tool to those of ASP LunarShader. We run on both tools a simple simulation for latitudes -88° to -90°, at surface level, for year 2010, for which LROC data is available for further validation. We run a six month simulation to speed up computation, but still using a 1 hour time step. We used a coarse gridding (spacing between the analysed points of 240 m) and a map with a resolution of 240 m per pixel. We compared the results in terms of accumulated illumination time, and verified that the five most illuminated points are the same across the tools (see Figure 15 and Table 4). This is the case for all points except point 5, with a distance between the two points returned by the different tools of only 690 m. For the five most illuminated points, we compared the accumulated illumination, i.e. the percentage of times at which the Sun is visible over the total duration of the simulation, and verified a good agreement, with a difference of always less than 4% of the total (see Table 4).

We generated a set of instantaneous illumination maps (snapshots) of the RoI's, containing greyscale values proportional to the visible Sun fraction simulated at moments in time when NAC images were taken[4]. We then registered by means of ground control points and compared the snapshots to the NAC images. Figure 16 and Figure 17 show a result of this comparison. In general, the results of the simulations agrees with the images, expect for the large area in the bottom left area of Figure 17. Small features are captured (for example the craters around ($X_{stereo}$, $Y_{stereo)}$ = (-10.5, -11.9) km in Figure 17), although some appear shifted. Clearly, features at a scale smaller than the LOLA resolution (here 40 m) are not captured (craters at ($X_{stereo}$, $Y_{stereo)}$ = (-11.1, -12.3) km). Note that the registration errors between the snapshots and the NAC images have been estimated in the range of 5 m (RMS), by systematically measuring on the NAC images the distance between the borders of the shadow in the images and the snapshot.

We also compared the LQCIP simulated with ESA and Astrium tools, for a limited number of RoI's, at different heights above the surface, using several short darkness filters and different griddings. The Astrium tool was used with a threshold of 50%, equivalent to the point source simulated by the Coverage Tool. The comparison showed that the general pattern of the areas, and therefore the size of the potential landing areas, are consistent (see Figure 18 for a sub-area of CR1, 2 m height, 60 hour darkness filter and 40 m gridding), although some inconsistencies exist locally, in terms of duration of the LQCIP for several pixels. Globally, a root mean square error of ~44.7 days has been observed for this case. These inconsistencies do not appear to be correlated with the terrain elevation or slope and are considered to be mainly related to the method of calculation of the horizon. In particular, the differences are considered to be due to the method used to sample the terrain data, which clearly produces larger differences in the terrain elevation for closer terrain (see Figure 19). It has also been noted that the differences increase when using a darkness filter of 0 hours or a height of 1.2 m, and decrease when moving to GR1, probably due to the flatter terrain. In general, there doesn't seem to be a trend in which one of the tools produces the worst case results.

The horizon generated with the ICAT was also compared to those reported in Mazarico et al. (2011a) showing a good match. Results from all tools were found to be consistent on a global scale (-88° to -90° latitudes, 240 m resolution) among each other and with results from literature, obtained both from LOLA-based simulations (Mazarico et al., 2011a) and from accumulation of LROC images (Figure 15 compares very well with results from Speyerer and Robinson, 2011).

## 3. Characterization of the hazard distributions

Landing hazards can exist at the sites identified by the illumination analyses. With the current lander design, hazards are defined as slopes steeper than 15°, roughness features higher than 50 cm and shadowed terrain. The slope is defined as the inclination relative to the local horizontal of the mean plane (see Figure 20). The mean plane is defined as the plane that best fits the terrain, on a baseline of approximately 10 m, which is the sum of the lander footprint (including the legs) and the control errors at touchdown. Roughness is defined as the deviation from the mean plane, and as such it is a property of each point of the terrain below the lander and can consist for example of small craters or boulders.

---

[4] Only a binary map was produced with ESA Coverage Tool, because it simulates the Sun as a point source. In addition, due to limitations of STK, the point at which the Sun visibility was simulated must be placed at 30 cm height to avoid artefacts in the computation of the horizon when using high resolution terrain.

A vast literature on terrain slope, terrain roughness, craters and boulder distributions exists. For example, the studies of Bart and Melosh (2005), Carrier at al. (1991), Hörz et al. (1991), Hutton (1969), Nagumo and Nakamura (2001), Rosenburg et al. (2011), Smith and West (1982), Wu and Moore (1980) provide extensive data that have been used in the preliminary design of ESA Lunar Lander. However these studies do not in most cases cover the areas of interest for the Lunar Lander mission and they use data resolutions of at least one order of magnitude coarser than those of interest here (from sub-metre scale to tens of metres). Therefore, extrapolation of the results therein is not sufficiently reliable to assess the hazard level of the potential landing sites.

The distributions of hazards at the locations of the Regions of Interest (RoI) identified by the illumination analyses are being assessed by independent studies carried out by Birkbeck College, Freie Universität Berlin and the Institute of Space System of the German Aerospace Centre (DLR). The objective of these studies is to assess the density of hazards over the total landing area, the compatibility of the hazard distribution with the HDA system and lander performances, mainly in terms of capability to divert (distance from nominal, unsafe point to the safe point), and ultimately the risk associated to the provisional landing sites. This assessment takes as much as possible into account the uncertainties in the data.

LOLA products are used to assess slopes on a long baseline. Lunar Reconnaissance Orbiter Camera (LROC) images are used to extract craters and boulders, visually and using computer tools, and shadows. The above data have been pre-selected (based e.g. on illumination conditions etc), processed (calibrated, projected etc) and integrated (including co-registration), using the Integrated Software for Imagers and Spectrometers (ISIS, 2011) and the Geographical information System software package ArcGIS (ArcGIS Desktop, 2011).

## 3.1 Slopes

Terrain slope can cause the lander to capsize if above the required 15°. For the exact determination of slopes on the lander footprint scale of ~5 m, none of the available laser altimetry datasets (e.g. LOLA) can qualify (see Section 2.2). It is however possible to extract slopes at scales of more than 5-10 m/pixel, and construct the slope distribution as a function of baselength at different scales, using raster-based (pre-gridded) and vector-based topography datasets.

Based on LOLA Gridded Data Records (GDR), we calculated the terrain slopes on 5 m grids, using the Slope function of the ArcGIS 3D Analyst extension (ArcGIS Desktop, 2011) using an 8+1 neighbourhood roving window. We then resampled the grid iteratively, from a resolution of 5-10 m to 40-60 m depending on the RoI, using cubic convolution interpolation, in order to calculate a best fit curve (average slopes as a function of baselength). We used an analysis area centred on the point with maximum terrain elevation and with a radius of 250 m, in order to allow for extraction of statistically significant data (except for CR1 and CR2, for which a radius of 150 m was used in order to avoid the steep slopes of the Shackleton crater interior). Figure 21 (top) shows the results of this analysis for the vicinity of the CR1 landing site. On the right, slope maps are shown, each using a different pixel scale that has been incorporated during cubic-convolution pixel resampling. Binning is the same for all subfigures and slope angles larger than 15° are shown in red.

Based on LOLA pulse measurements (Experimental Data Records), we calculated slopes on a Triangulated Irregular Network (TIN) basis, derived from the Delauney triangulation of the measurements. Figure 21 (bottom) shows the point distribution, the TIN based on Delaunay triangulation and the derived slope map for CR1. Analysing the TIN-based slope maps and comparing these with the GDR-based slope maps at different baselengths, it appears that areas with 5 m slopes higher than 15 degrees are correlated with the areas with higher data density, which is possibly an effect of data error (Section 2.2), whereas slopes at 10 m baselength or longer appear much more shallow. The distribution of slopes as a function of baselength, with estimated errors, is given in Figure 22 for all sites, showing that slopes level off for baselengths longer than 20 m, and that the Shackleton rim sites exhibits steeper slopes than the other sites even at long baselengths.

Owing to the sparseness and errors of the LOLA points (Section 2.2), and to the requirement that several points occur within each pixel to derive a reliable slope estimate, we repeated the slope analysis over the entire RoI's, using LOLA GDR maps at 40 m/pixel resolution. At this scale, within all the RoI's areas exist with average slopes <15°. This conclusion appears robust for sites with good LOLA coverage (SR1, SR2, CR1, SV1), but more data is required for the other sites (GR1, MP1, MP2, LP1), which are not yet adequately sampled on this scale. Therefore other methods of slope determination (e.g. slopes derived from stereo LROC imagery) are required to derive slopes on the scale of the lander itself. In the absence of data with metre scale resolution, we provisionally conclude that slopes at baselines smaller than 40 m are dominated by craters.

## 3.2 Craters

Craters are the most common feature of the lunar terrain, and represent a hazard for landing safety as well as for surface mission success. Assuming a simple, young crater shape model (Melosh, 1989), craters can be considered as roughness or slope hazard depending on their size. Craters with diameters between ~4 m and ~10 m present a roughness hazard, because the combined crater depth (20% of the diameter) and rim height (4% of the diameter) would, in a worst case in terms of position of the landing pads (one inside and one outside of the crater), result in a roughness higher than the 50 cm limit. Craters larger than ~10 m present a slope hazard, due to the internal slope of the crater. Note that a keep-out zone of half the lander footprint shall be kept around the craters. In addition, considering the low Sun elevation at polar sites, craters are likely to cause significant shadowing (~14 days), depending on the slope of the underlying terrain, and therefore constitute a potential risk for the success of the surface mission phase. It shall be noted however that assuming that all craters are young is conservative, as most craters in the south polar region can be expected to be aged, in which case they have internal slopes compatible with the lander touchdown stability requirement (≤15 °) and lower rims[5].

Crater-size frequency data were derived using manual measurements within an integrated GIS environment (ArcMAP and the integrated Crater Helper Tool plug-in, ArcGIS Desktop, 2011).

---

[5] A small aged lunar crater is expected to have a depth to diameter ratio close to 0.1, and a rim height of about 1% of the diameter (Hutton, 1969). The general assumptions that craters are aged in the south polar region is also supported by the observed crater size-frequency distributions for all sites (except for LP1, see below).

Measurements were conducted by digitizing crater rims as defined by shadowed areas. The misdetection is on the sub-pixel level and a sophisticated shifting centre of map-projection eliminates any errors caused by map distortions (Kneissl et al., 2011). Measurements were carried out for SR1, MP1, MP2 and LP1 on large areas (up to ~50 km$^2$) around the potential landing sites (see Figure 23 to Figure 26), down to a scale of ~2 meters in crater diameter, and on smaller windows within the larger areas, down to the limit of detectability (1.3 m best case). Measurements were carried out on small windows only for SR2 and CR1 (0.2 km$^2$).

The detection of impact craters is in principle straightforward but their digitization relies on some experience in this field in particular when dealing with different illumination conditions. In order to avoid misdetections, craters have been counted on multiple image mosaics covering each landing site at varying illumination conditions. In particular for MP1 and MP2 this procedure was helpful to improve the detection.

For all measured areas, the distribution is close to the equilibrium distribution, as it is expected for the lunar surface (Hörz et al., 1991). LP-1 shows a significant kink in the distribution which may have been caused by a widespread resurfacing event such as more recent ejecta covering the potential landing sites. After that event, the curve approaches the equilibrium distribution again, and as a consequence at LP-1 the number of craters in the 10-100 m range is slightly reduced. Although the distribution of impact craters closely follows expectations, local fields of secondaries should be localized. By linking surface ages derived from crater-size frequency statistics using appropriate model functions (Michael and Neukum, 2010) it should be possible to determine some age constraints for the weathering and disintegration of rocks and boulders at each landing site.

### 3.2.1 Dependence on illumination conditions

Because of the extreme illumination conditions at the pole, where the Sun can never rise more than few degrees above the horizon, it is important to determine whether the crater (and boulder) statistics are sensitive to the low illumination angle. We have examined this by comparing the crater (and boulder) statistics at a lower latitude site (specifically that close to the Apollo 16 landing site) for which a much wider range of illumination conditions are represented in the LROC data archive. We have performed crater counts in exactly the same areas of three images of the Apollo 16 landing site. These three images, and their solar incidence angles, are given in Table 5. Figure 27 shows the cumulative crater size-frequency distributions as a function of illumination angle.

It is clear from this analysis that at the largest incidence angle (87.78 degrees; M132732855R) a large number of the smallest craters (<4 m diameter) have apparently been missed. This is important as this is the illumination angle which best approximates the best polar incidence angles. The implication is that counting craters at the poles in LROC images may underestimate the number of small (<4 m) craters which are actually present, by perhaps as much as a factor of two. This is precisely the size range of most importance for the Lunar Lander, and below the size range where LOLA data will be able to help in the identification of craters (as each LOLA beam only has a 5 m diameter footprint on the surface).

## 3.3 Boulders

Boulders can be common on the lunar surface and present a potential landing hazard, because they can damage parts of the lander structure at touchdown or by contributing to slope instability, in case the lander touches down with a landing pad on a boulder. A detailed boulder analysis for the areas centred around the potential landing sites has been performed. Boulders were visually identified using image mosaics comprising data taken under varying illumination conditions (Figure 28). The pixel scale of these mosaics is between 0.6-1.0 m per pixel, thus they allow identifying boulders in the range down to 1-2 meters, depending on the shadows (see below).

We used ArcMAP (ArcGIS Desktop, 2011) and the ArcGIS extension CraterTool (Kneissl et al., 2011) to locate and estimate the sizes of boulders. Boulders were digitised by using a circular vector representation. This method provides a common measure for statistically representing boulders while being a feasible simplification, due to the fact that most boulders are below a size that permits outlining a polygonal feature. Boulders do in fact resemble small inverted impact craters whose detection is relatively straightforward (see Figure 28). Due to their small size, radii are usually slightly overestimated, as our personal experience with impact-crater measurements and multi-scale comparisons has shown. The shadow is important for estimating the extent of a boulder: if the feature is highly irregular the radius is either under- or overestimated. As lunar boulders are relatively regular in shape, i.e. the average ratio of maximum to minimum diameter is considered to be close to unity (Hovland and Mitchell, 1973), we do not expect any systematic measurement trends.

For all broader regions that were investigated, no homogeneous distribution of boulders could be identified. Boulders appear as clusters with a high density of tens to hundreds of boulders per cluster field (highly variable in size). For the 9 km$^2$ area around MP2, 62 boulders were detected. The largest boulder has a diameter of 11.1 m, 50% of all boulders are larger than 4 m in diameter. The 4 km by 2.5 km area on the Shackleton crater rim around SR1 and SR2 shows 820 boulders; the largest one has a size of 18 m in diameter, the smallest one 1.3 m. The size frequency distribution curves for the entire MP2 and SR1/2 areas (Figure 29) show characteristic kinks at the 6 to 10 m diameter level respectively, which can be explained by differences of the distributions inside each cluster. The distribution across clusters, however, compares closely to the results published in literature (Hutton, 1969, Smith and West, 1982). Note that, although separating the clusters would be more appropriate for the construction of size frequency distributions, it was important for the hazard analysis to have an estimate of the upper limit of the number of boulders of a given size, which we obtained by merging the distributions.

In the immediate vicinity of the landing sites only a few boulders were identified in the detectable range (see Table 6). The lack of a statistically meaningful number of samples implies that nothing can be inferred for boulders below the detectable size. However, it is unlikely that a site is populated with boulder below the detectable size when no boulder can be detected and no significant cluster exists in the vicinity. This assumption is deemed reasonable for this study but needs further validation. Note that based on the analysis of the effects of the Sun incidence on boulder detection (Section 3.3), we are confident that the low areal coverage of boulders is real and is not an artefact of the illumination conditions.

### 3.3.1 Dependence on illumination conditions

As for the crater counting discussed in Section 3.2, it is important to study the effects of the oblique lighting conditions on the boulder distributions obtained. We therefor counted the boulders in the same areas in the vicinity of the Apollo 16 landing site discussed in Section 3.2 at a range of illumination conditions. The results are given in Table 7). Clearly the size distribution is different between the Apollo 16 and polar landing sites, even though the total area covered by boulders is similar. Note that this exercise clearly shows that we can identify boulders as small as 1 m across, and that the total number identified does not depend strongly on the illumination angle. The most obvious trend in Table 6.2 is that boulders appear smaller at low angles of incidence (i.e. more vertical illumination), and it may be that the sizes of boulders at the polar landing sites are somewhat overestimated.

### 3.4 Automatic terrain feature detection

Development of software for automatic detection of hazardous terrain features has been started at DLR Institute of Space Systems, in order to speed-up the feature detection process and to collect extensive statistics. The underlying assumption, which reinforces engineering safety, is that any clearly visible and discernible terrain feature is regarded as potentially harmful to the landing system, regardless of its real nature. Terrain features refer therefore to all potentially hazardous objects undulating the terrain surface, and include boulders, small craters, scarps etc.

The automatic feature detection follows a two-step approach applying image processing techniques to the LRO NAC image data. The first step makes use of texture filters (Bajracharya, 2002) to segment the image and outline potential features, without discriminating between boulders and craters. Further morphological operations (image reconstruction and opening) are applied to extract the resulting clusters of pixel. The second processing step classifies the potential features as crater or boulder by exploiting their characteristic bright/dark pattern. This is done by means of image correlation between the respective pixel cluster and templates representing crater and boulder illumination pattern (Stepinski et al., 2012, Kim et al., 2005). The algorithmic concepts described above have been implemented in the MATLAB environment (MATLAB, 2010) and, whilst not new in essence, their implementation and the parameter settings have been tuned to meet the extreme illumination conditions in the lunar polar region.

Preliminary results show a good agreement of the automatic feature extraction with the visual boulder and crater detection, in terms of correct identification and classification of the features and in terms of the derived size-frequency distributions. As an example, Table 8 shows several quality measures of the automatic identification and classification performances for boulders and craters, derived using as reference data a sub-set of the results of visual detection for the SR1 Region of Interest, reported in Sections 3.2 and 3.3. True Positives (TP) are features (craters or boulders) that are detected and correctly classified; False Negatives (FN) are features that have not been detected; False Positive (FP) are features that have either been incorrectly detected (for example a non-feature undulation of the terrain) or correctly detected but incorrectly classified (for example a crater classified as a boulder or vice-versa). The following measures are also used:

    Detection Percentage = 100·TP/(TP+FN)

    Branching Factor = FP/TP

The Detection Percentage (correctly detected features over total number) is comparable with state of the art performances for craters (Stepinski et al., 2012, Kim et al., 2005). The Branching Factor instead indicates that, for every 10 true craters, 3 are false alarms, and that for each boulder almost 9 false identifications are counted. Visual inspection indicates that false positives for boulder are caused by small, decayed craters whose bowl is too shallow to appear as a sharp dark/bright pattern, such as that of a fresh crater, while the remains of its rim are illuminated. The algorithm detects these craters as boulders, thus the majority of false positives for boulders equals the false negatives for craters. Whilst certainly unsuitable for scientific data interpretation, this bias from decayed craters to boulders has the net effect of providing a conservative assessment of the total terrain hazard density.

Figure 30 shows a comparison between the cumulative number per unit area of the craters automatically extracted in the SR1 RoI with the cumulative number per unit area of the craters visually detected in the same RoI. The curves generally show a good agreement, although the number of missed detections increases below 10 m (see explanation above). The plot stops at 20 m because this is the upper limit deliberately set for the feature size in the automatic extraction routine in order to avoid false alarms from large non-crater shadowed depressions (which are accounted as shadow hazards, see Section 3.5).

### 3.5 Shadows

Shadow hazard assessment is currently based on the analyses of LROC NAC images taken at times at which the illumination conditions (Sun azimuth and elevation angles) are equivalent to those expected at the estimated landing date and time. A threshold is applied to the Digital Number (DN) of the 8-bit NAC images, and pixels with DN lower than the threshold are considered in shadow and hazardous. A value of 5% of the maximum DN value has been selected as threshold. The image histogram shows in fact a characteristic distribution of the intensity values (Figure 31), where the intensity range from 0 to 5% of the maximum DN reveals a narrow peak to the general distribution. Assuming this as shadow footprint leads to a robust criteria to select the threshold.

Large shadowed areas, due to the low illumination angles, appears to be the most significant hazard component when compared to all other hazards analysed. However, large illuminated areas have been also observed in all the analysed images, still providing sufficient area for landing.

## 4. Future work

The site characterisation work is being currently performed for landing sites identified as having the most favourable illumination conditions. We foresee to repeat the illumination analyses using the latest LOLA products release and all three tools. Further validation of the illumination tools and development of error models will also continue in parallel. Refined and systematic analyses of all the sites, as presented here, will be conducted.

In order to reduce the uncertainties in the illumination simulations and to improve knowledge of slopes at small scales (2-10 m) and possibly roughness, we intend to produce Digital Elevation Models of the potential landing sites based on LROC Narrow Angle Camera stereo images, and to use those in the illumination simulations and in the generation of slope and roughness maps. DEM production at the poles is challenging due to poor illumination conditions but preliminary results show that this is feasible. In addition, increasing the resolution of the terrain models in the illumination tools may require some adjustments and tuning of the tools.

More extensive work will also be performed on crater size-frequency distributions and on crater and boulder modelling for all sites. Shadow hazard distributions will also be modelled using dedicated simulations. We will continue the development and validation of the tools for automatic boulder and crater detection. We will also finalise the framework for the combination of the hazard distribution into risk maps, also taking into account a measure of confidence based on the errors associated to the input data. Detailed models of the landing sites will be produced and used in end-to-end landing simulations, in order to validate and verify in a realistic environment the performances of the Hazard Detection and Avoidance system and the landing legs dynamics.

In view of future mission and system design and development phases, a complete landing site certification and selection process will be put in place. This includes setting up a framework for the certification of the input data, developing and validating all the missing simulation tools and models, generating a detailed landing site environment specification and deriving an accurate landing site risk assessment, which takes into account all the uncertainties in the input data.

## Acknowledgements

The authors would like to acknowledge the LOLA team at GSFC and MIT and the Institut für Geodäsie und Geoinformationstechnik at Technical University of Berlin for the advice on LOLA data, as well as the ESA and industrial Lunar Lander teams for their support in the conception, development and interpretation of the results of the analyses.

# Figures

Figure 1. Orientation of Lunar North Pole, Lunar Orbit Pole and Earth North Pole relative to the Ecliptic North pole, and position of the Moon at lunar summer solstice, winter solstice and equinoxes.

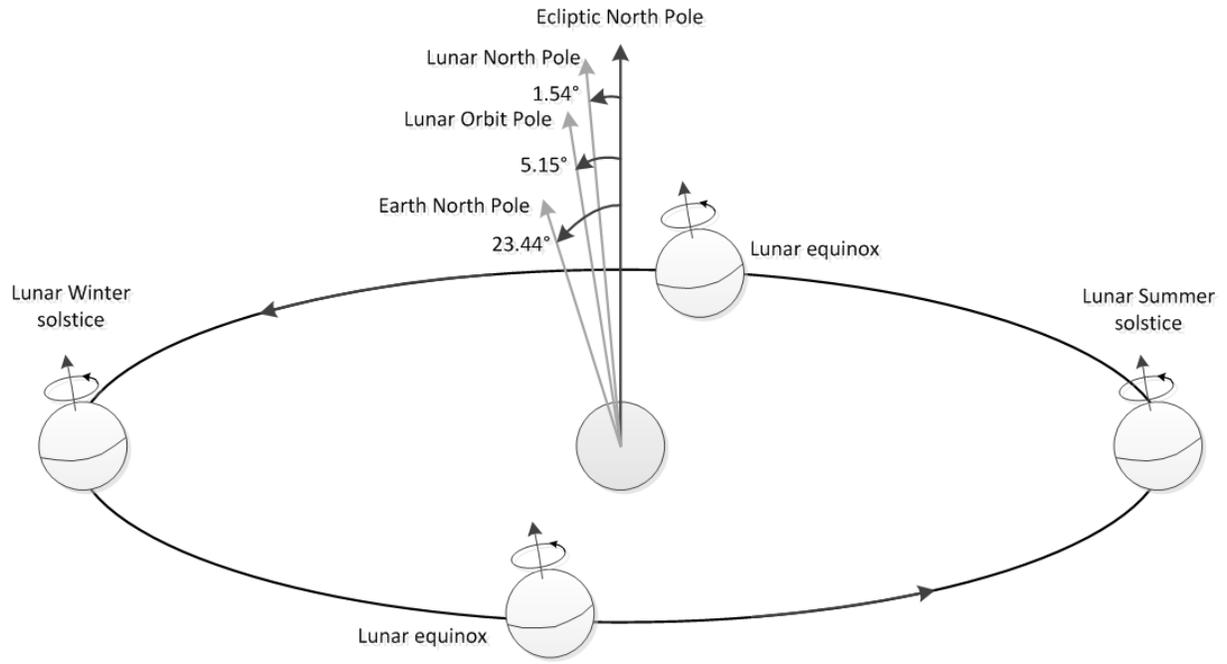

Figure 2. Local horizon (bold black line) and path of the Sun centre over one year (colour lines), as seen from a location at around -89.5° latitude and 222° longitude.

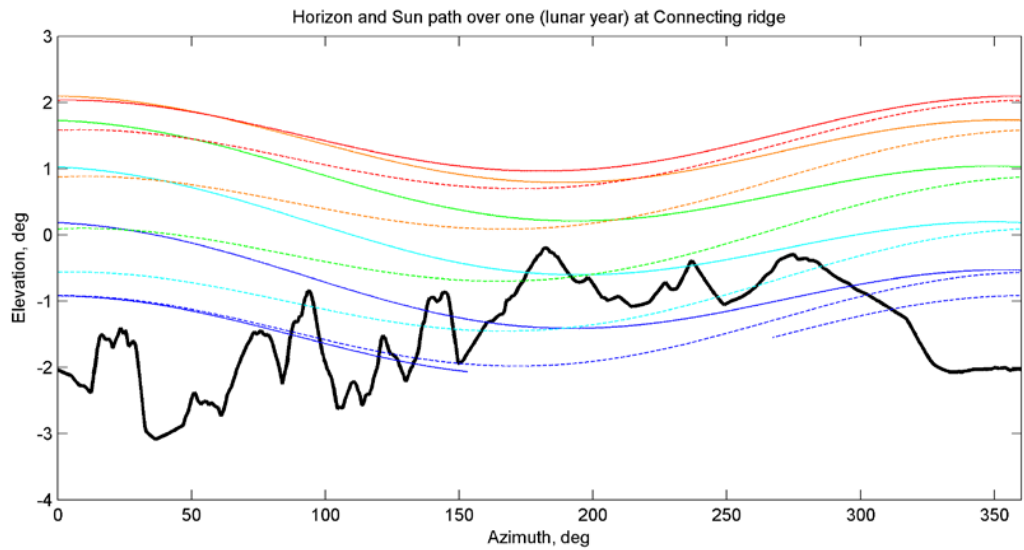

Figure 3. (a) Time history of the visible Sun fraction, for 12 synodic periods (~29.53 days) starting from lunar southern winter solstice (defined as the epoch of maximum sub-solar latitude) of 2018 (October 22, 2018); (b) Pattern of illumination and darkness periods after thresholding to 50% of Sun fraction; (c) Pattern after filtering of illumination periods shorter than 10 hours; (d) Pattern after filtering of darkness periods shorter than 60 hours. The Longest Quasi-Continuous Illumination Period is the longest illumination period in (d).

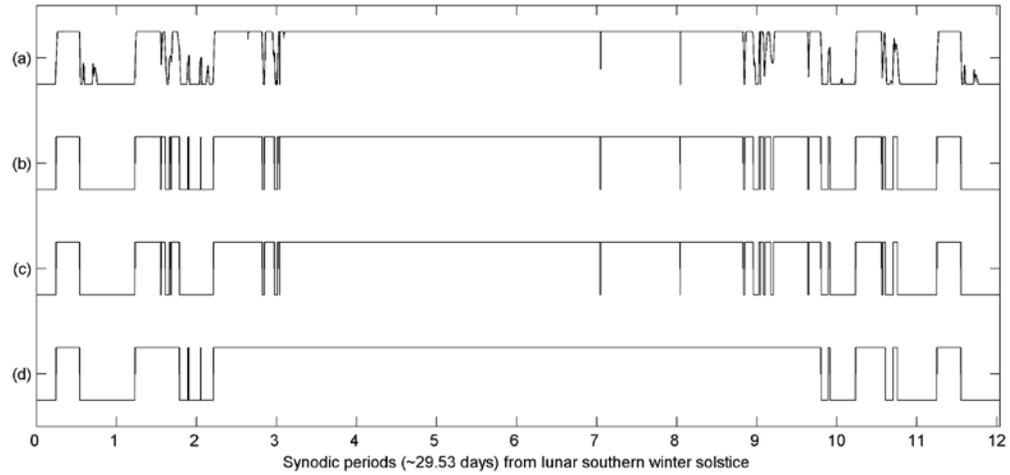

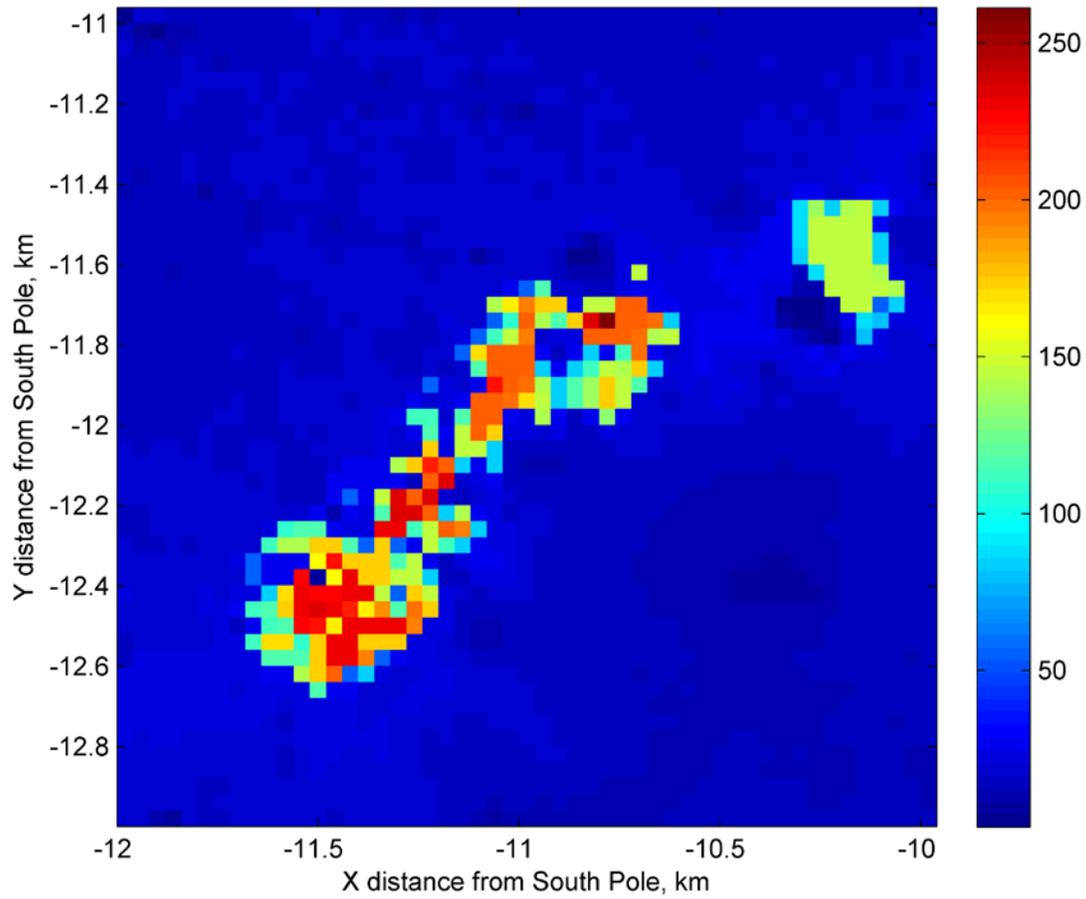

Figure 4. LQCIP map for the Connecting Ridge RoI, reporting the LQCIP duration, color-coded in days, for 2 m height above the surface, a filter for short darkness periods of 60 hours, a filter for short illumination periods of 10 hours, and year 2019. The spacing is 40 m. X, Y coordinates are in polar stereographic projection.

Figure 5. Map of the South Polar Region based on a LOLA DEM, with the locations of the Regions of Interest analysed in this study. Isolines (circles) are every degree of latitude.

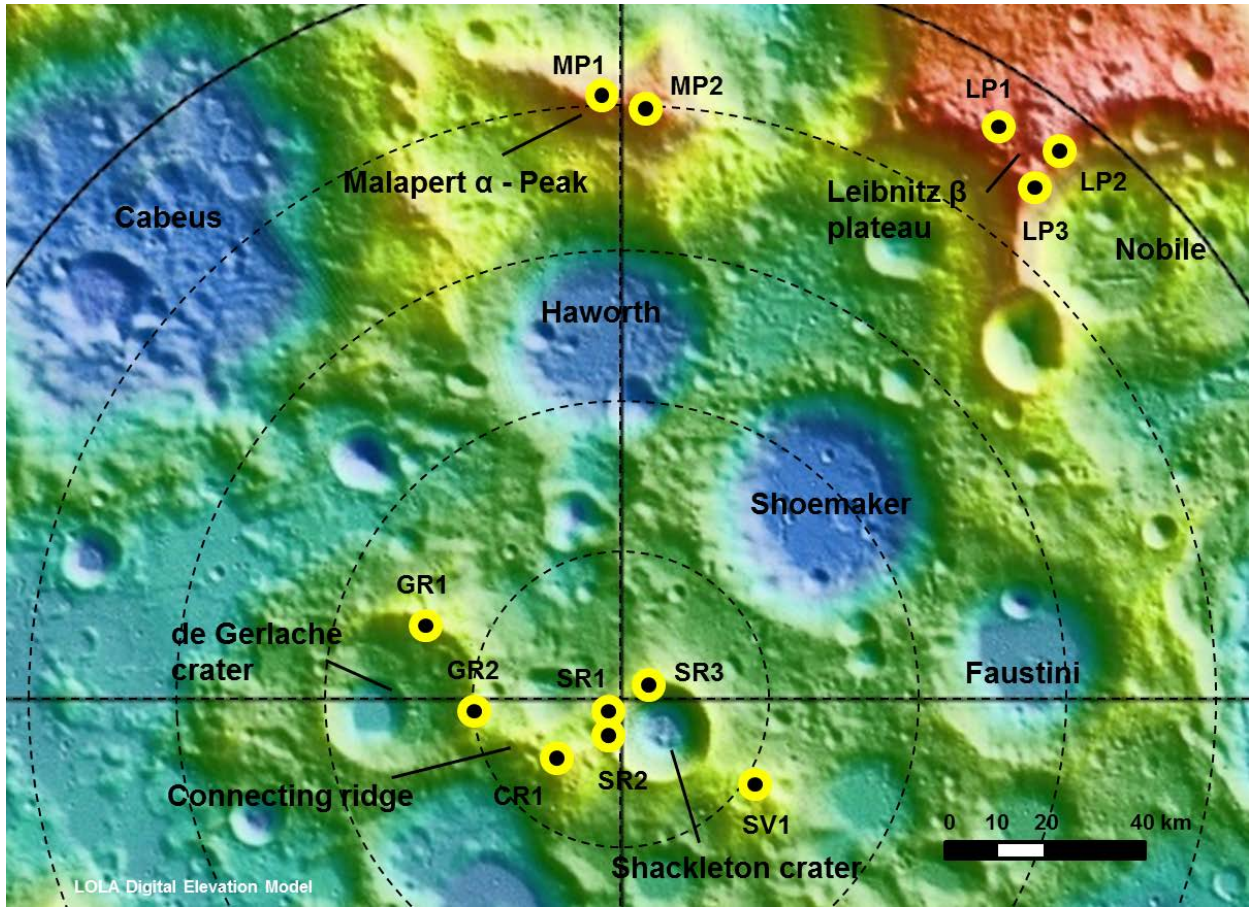

Figure 6. Computation in the Illumination and Communication Analysis Tool (Astrium) of the horizon elevation for an arbitrary azimuth. The horizon elevation at the Sun azimuth, corresponding to the current simulation time, is computed by linear interpolation of the elevation values at the surrounding points of the horizon. The figure also shows the visible Sun fraction computation by evaluation of the vertical fraction above the computed horizon elevation (area dashed with red lines, from top-right to bottom-left). The difference with the true visible fraction (area dashed with blue lines, from top-left to bottom-right) has been found to be negligible (0.4% of the visible fraction for a 1-year simulation, with very few cases above 1% difference), in view of a significant computational gain.

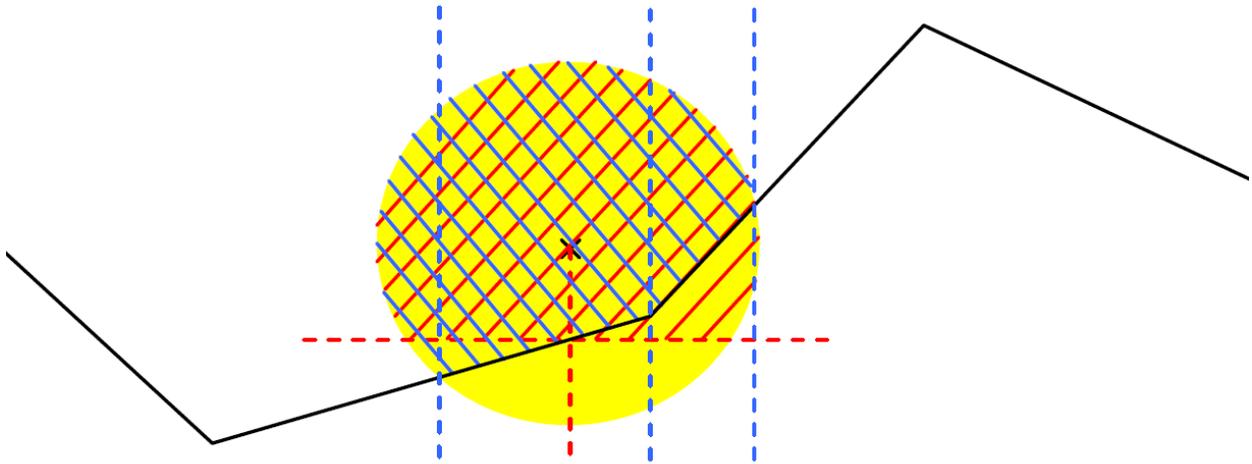

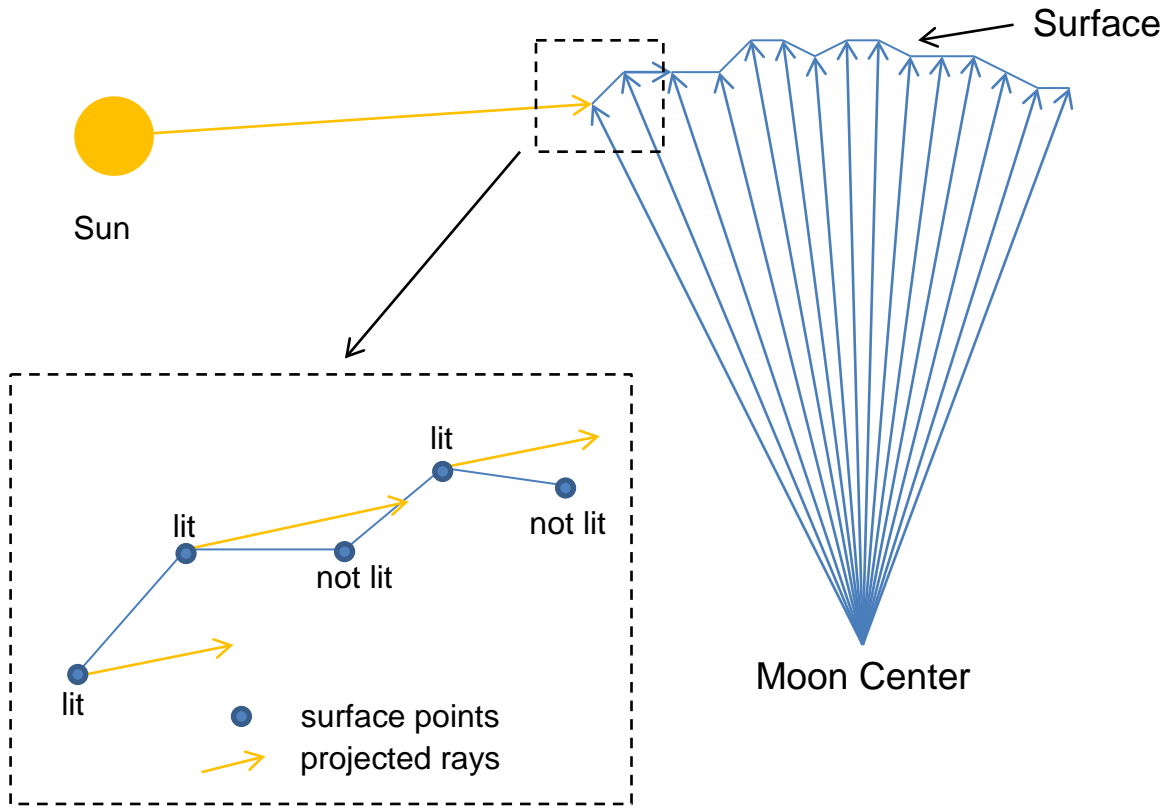

Figure 7. Projected light rays along a terrain path within a single plane and containing the Sun, the centre of the Moon, and an ordered set of surface points.

Figure 8. Hillshade of the LOLA DEM for the Connecting ridge area, superimposed to the individual LOLA measurements, colourised by track number. An example LQCIP map for the RoI is also given for reference. X, Y axes are in polar stereographic projection, in metres.

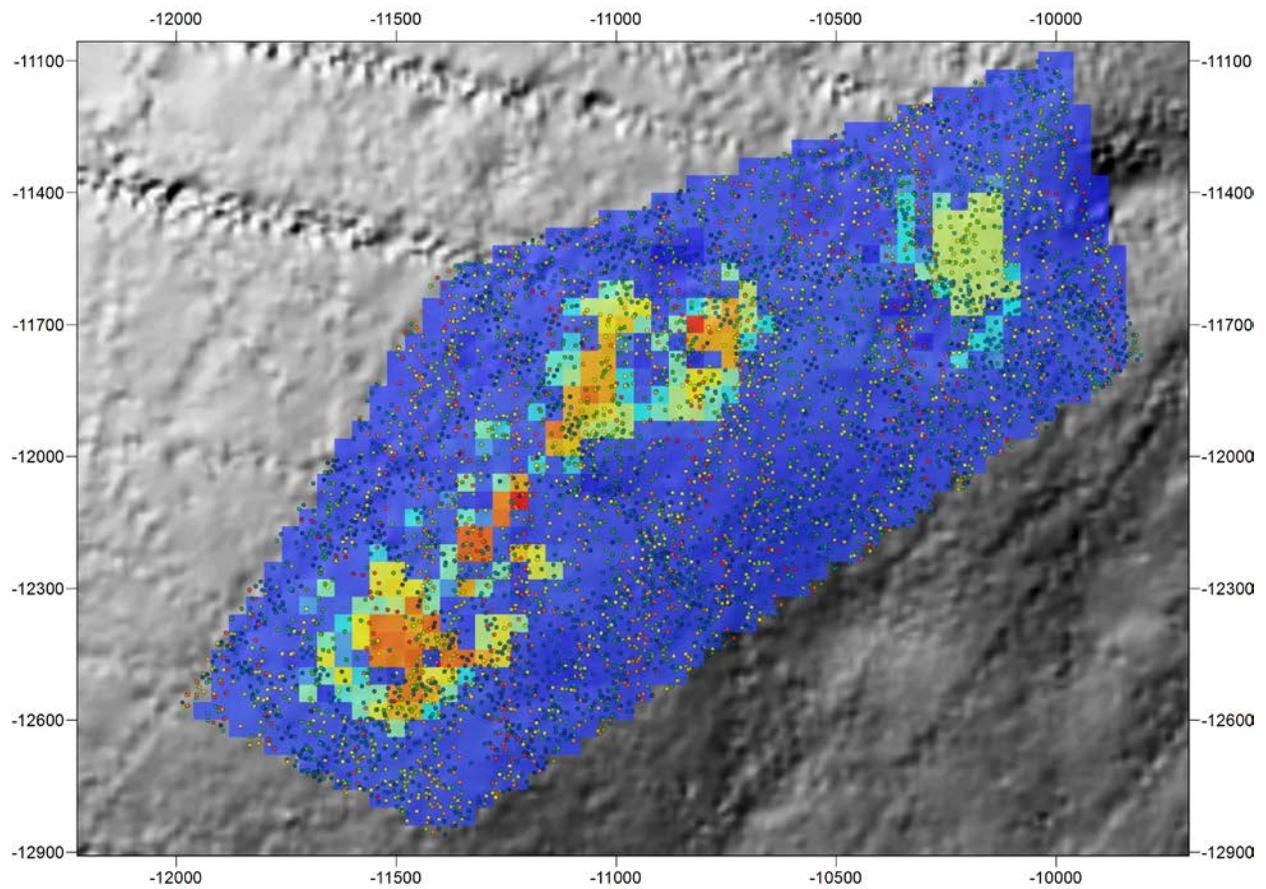

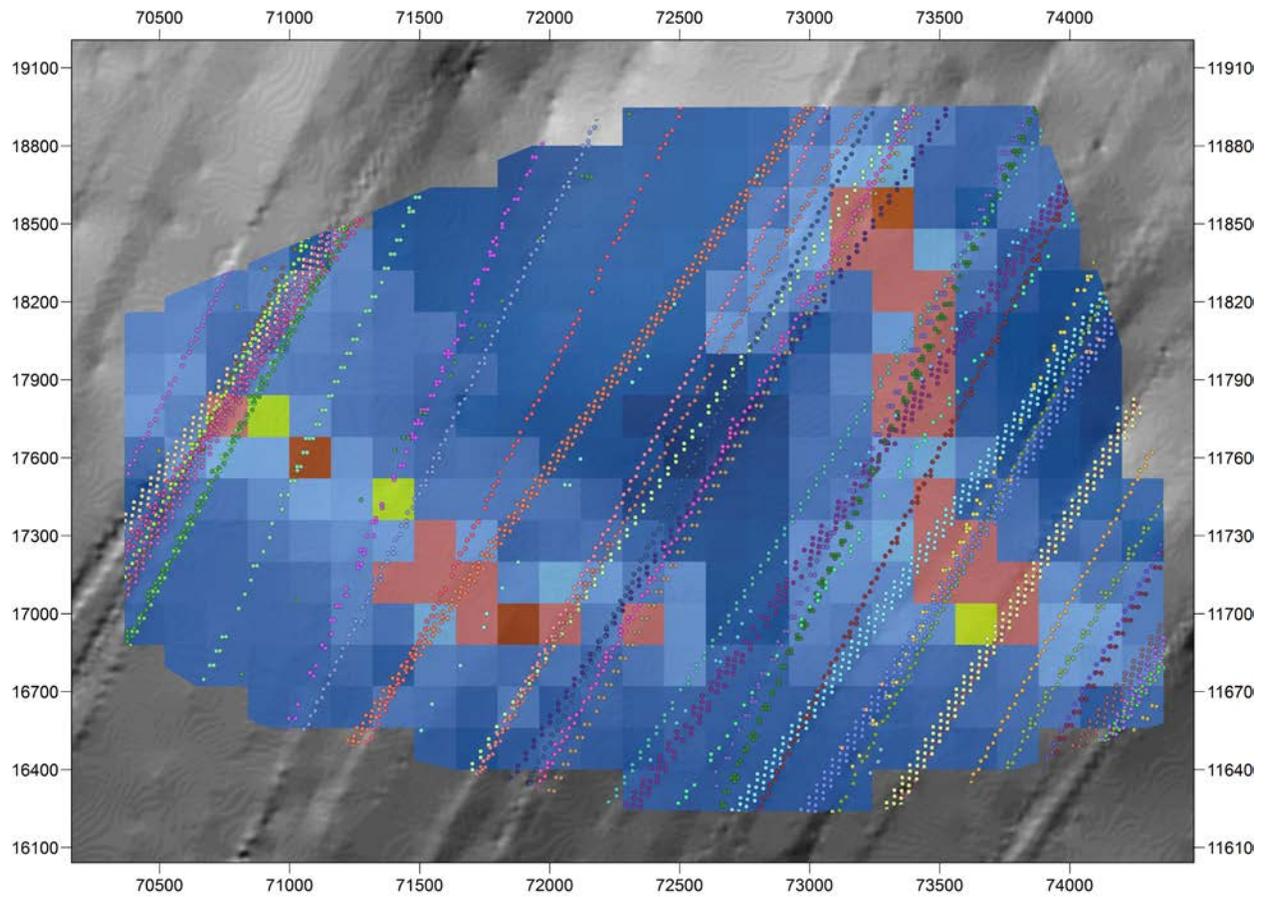

Figure 9. Hillshade of the LOLA DEM for the Leibnitz Plateau area, superimposed to the individual LOLA measurements, colourised by track number, and to an example LQCIP map (given for reference). X, Y axes are in polar stereographic projection (m).

Figure 10. LQCIP map for the six best RoI's, with RoI name and geodetic coordinates (in the Mean Earth/Polar Axis reference system). Colour-code is in days, height above the surface is 2 m, filter for short darkness periods is 60 hours, filter for short illumination periods is 10 hours, and year is 2019. The spacing is 40 m. X, Y coordinates are in polar stereographic projection.

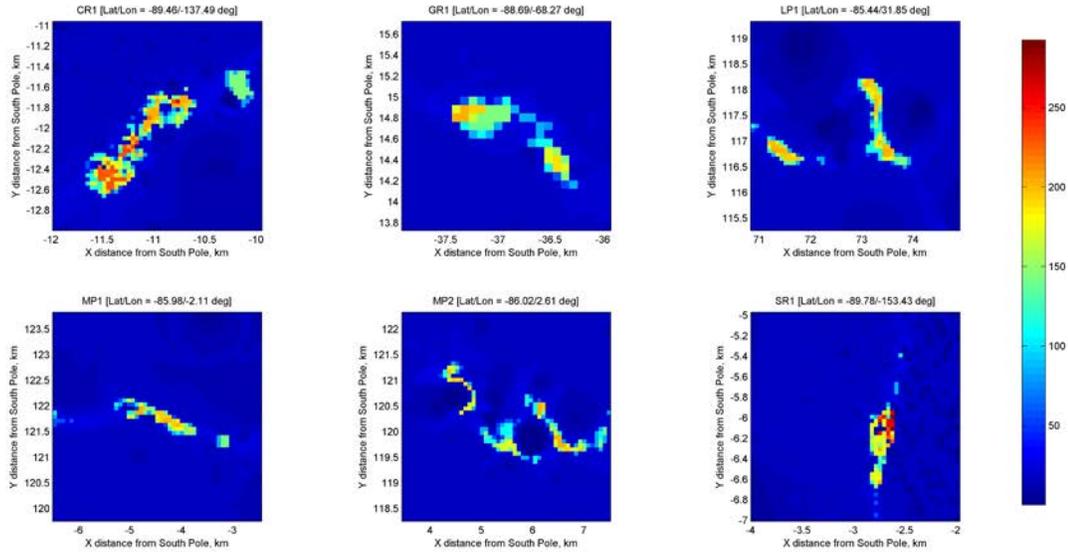

Figure 11. Map showing the LQCIP duration (in days) at the Connecting Ridge, simulated for year 2019, for different values of height (left to right) and short darkness filters (top to bottom). Grid spacing is 40 m. Coordinates are in polar stereographic projection.

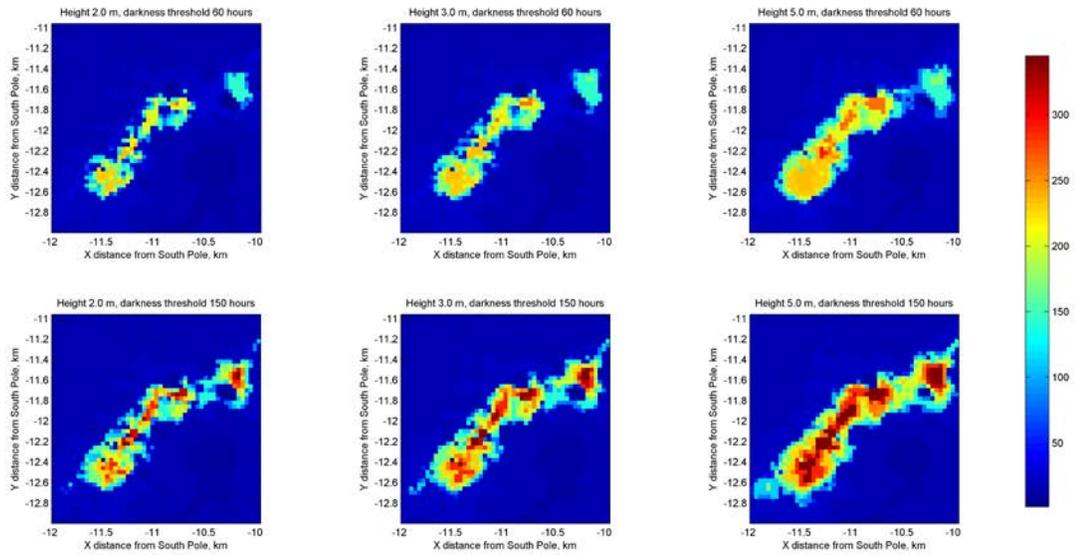

Figure 12. LQCIP maps for the Connecting Ridge, simulated for year 2019, for different values of the darkness/illumination threshold (left to right) and darkness filter (top to bottom). Grid spacing is 40 m. Coordinates are in polar stereographic projection. It can be seen that for both values of the darkness filter, the extent of the area is practically not affected in almost all cases, while the duration of the LQCIP decreases slightly (up to ~3% difference) when going from 25% threshold to 75%.

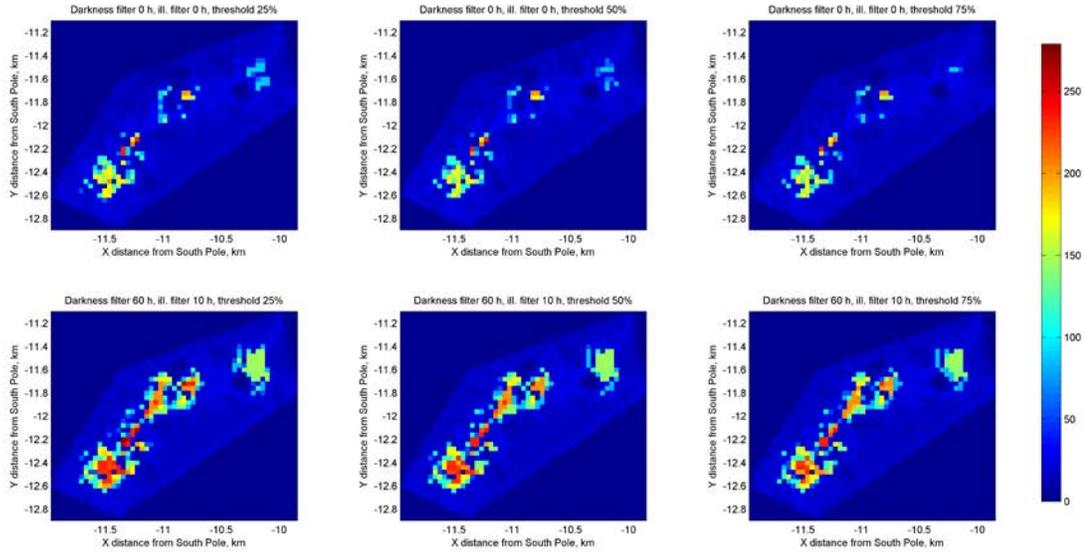

Figure 13. Cut-out of NAC image M112490681RE, with squares identifying terrain features which are always in shadow and correspond to gaps in the LQCIP maps (points with LQCIP duration shorter than 1 month).

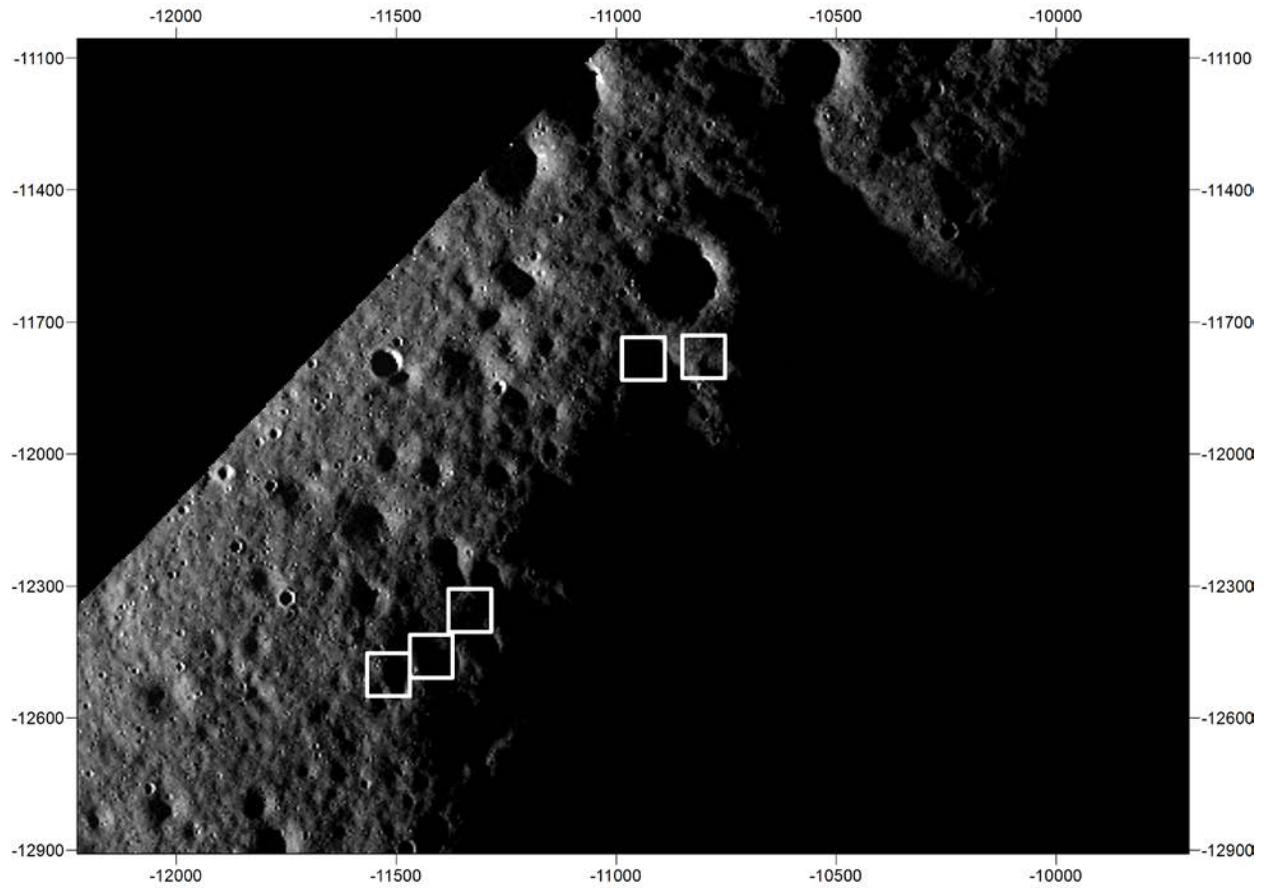

Figure 14. Cut-out of NAC image M118429312LE, with squares identifying terrain features which are always in shadow and correspond to gaps in the LQCIP maps (points with LQCIP duration shorter than 1 month).

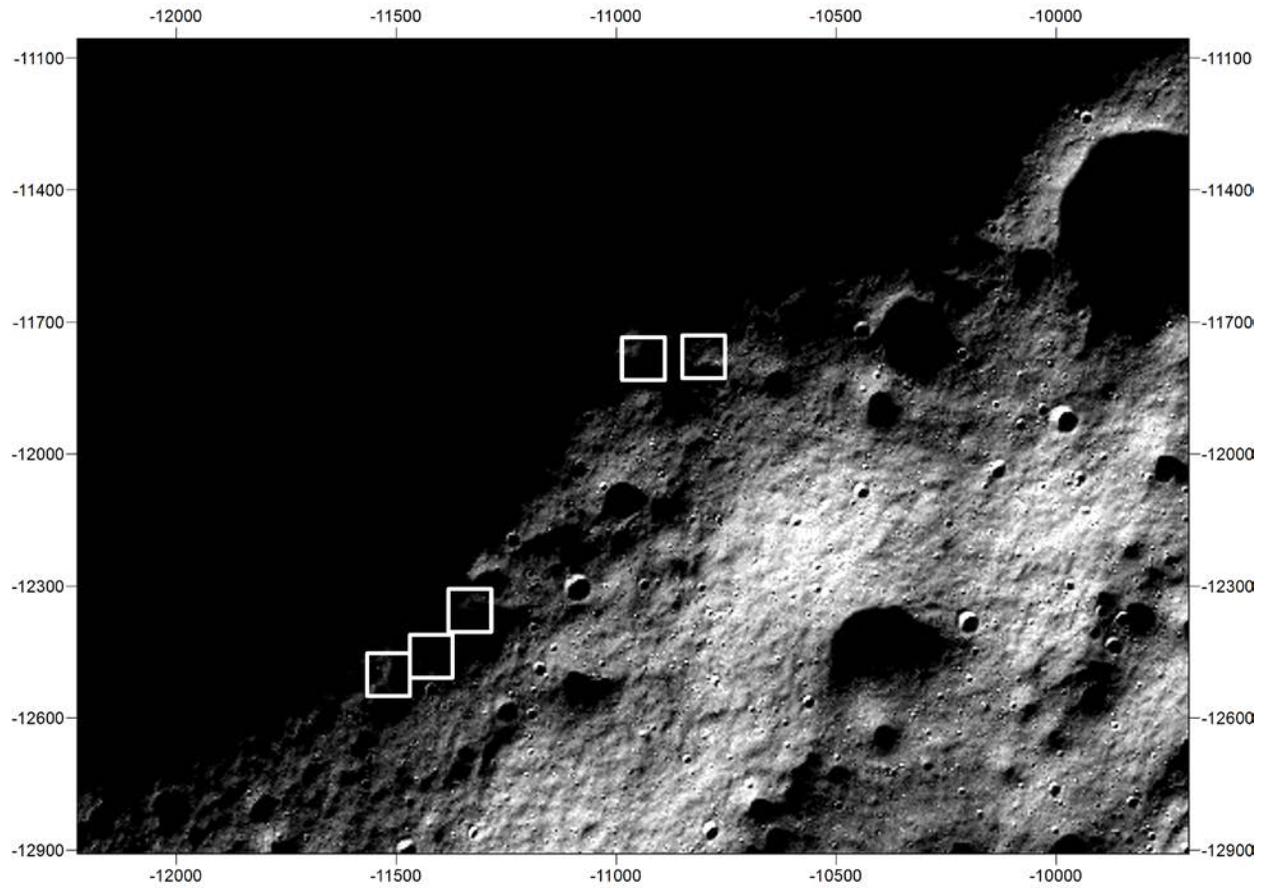

Figure 15. Map of accumulated illumination for the area 120 km by 120 km around the South Pole, as simulated with ESA Coverage Tool. Accumulated illumination is the percentage of time for which each pixel is illuminated, for period 2010/03/31 12:00 to 2010/09/24 00:00 (the simulation was run with a 1 hour time-step). Black is 0%, white is 100% of illumination. The squares indicate the 5 points in the region that receive the most illumination in the analysed period.

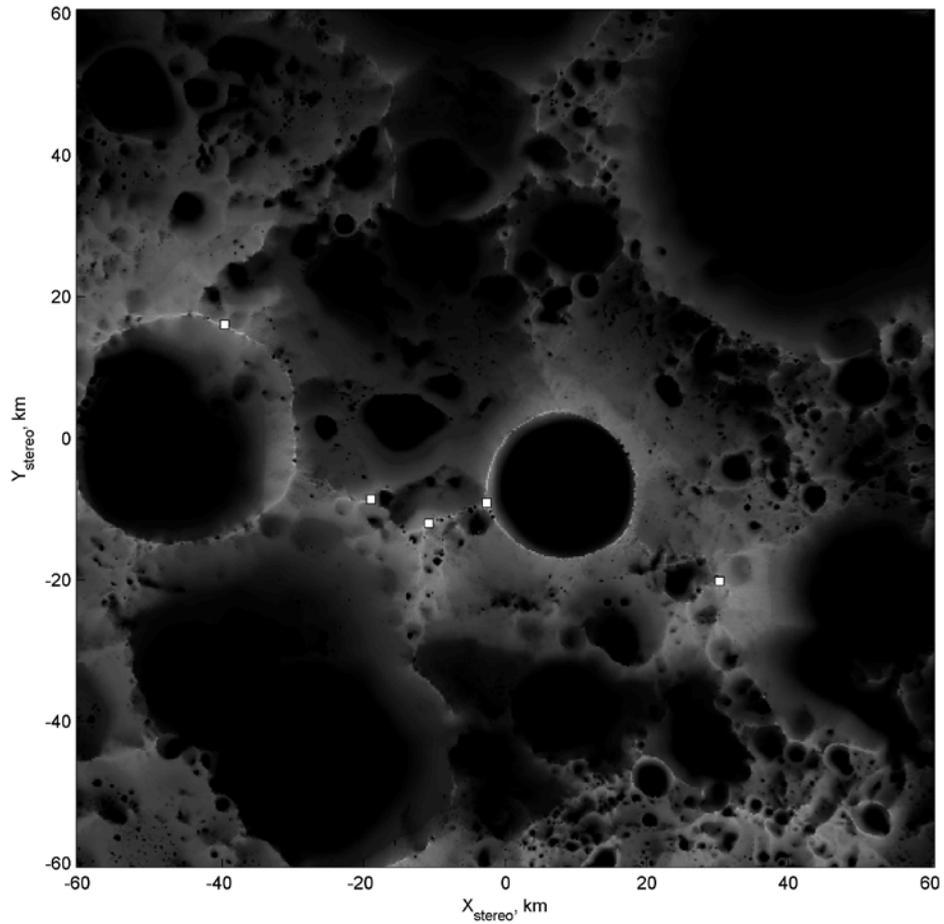

Figure 16. Cut-out of NAC image M112490681RE, with an overlay showing the Sun visibility conditions within the RoI's simulated with ESA Coverage Tool at that moment in time when the NAC was taken. Grid spacing and DEM resolution are 40 m, coordinates are in polar stereographic projection.

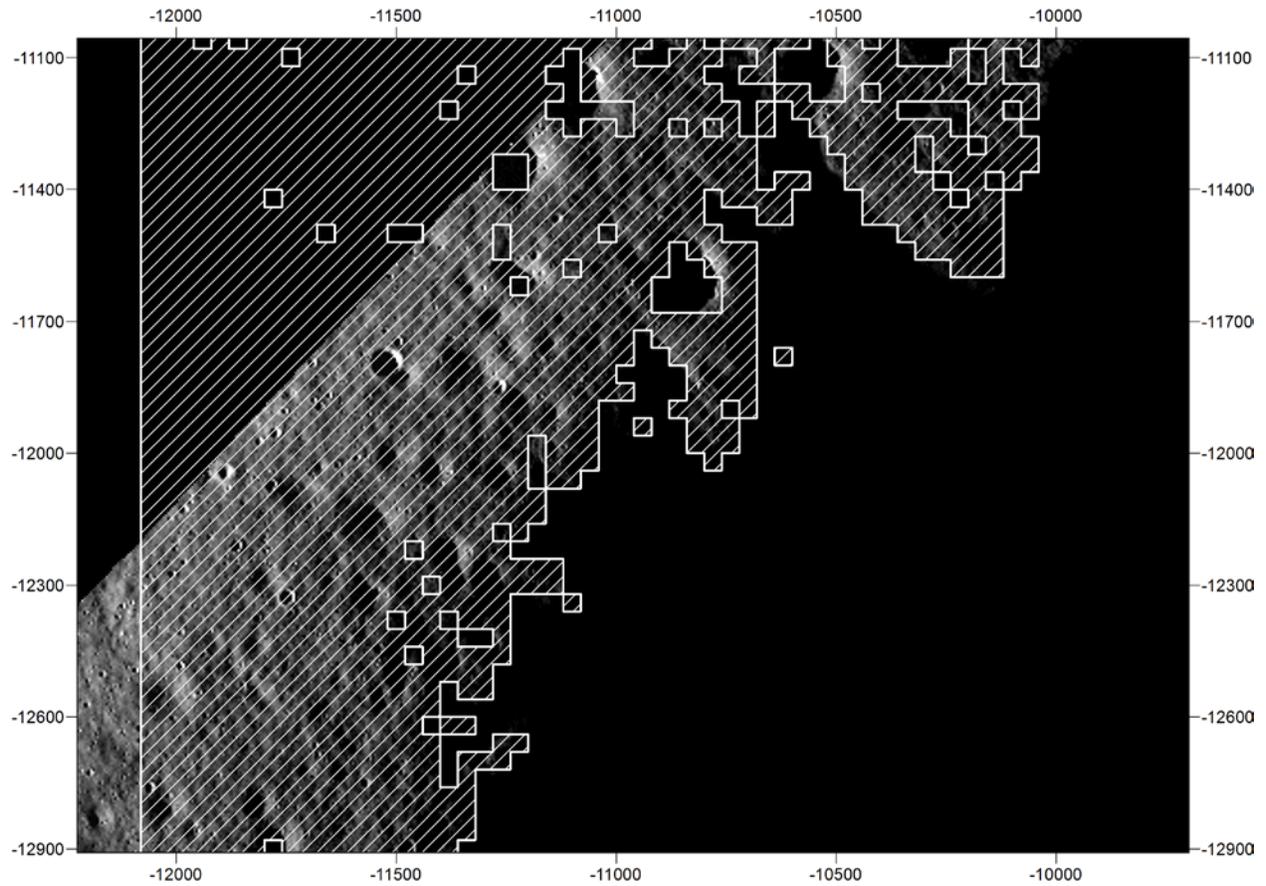

Figure 17. Cut-out of NAC image M118687162RE, with an overlay showing the Sun visibility conditions within the RoI's simulated with ESA Coverage Tool at that moment in time when the NAC was taken. Grid spacing and DEM resolution are 40 m, coordinates are in polar stereographic projection.

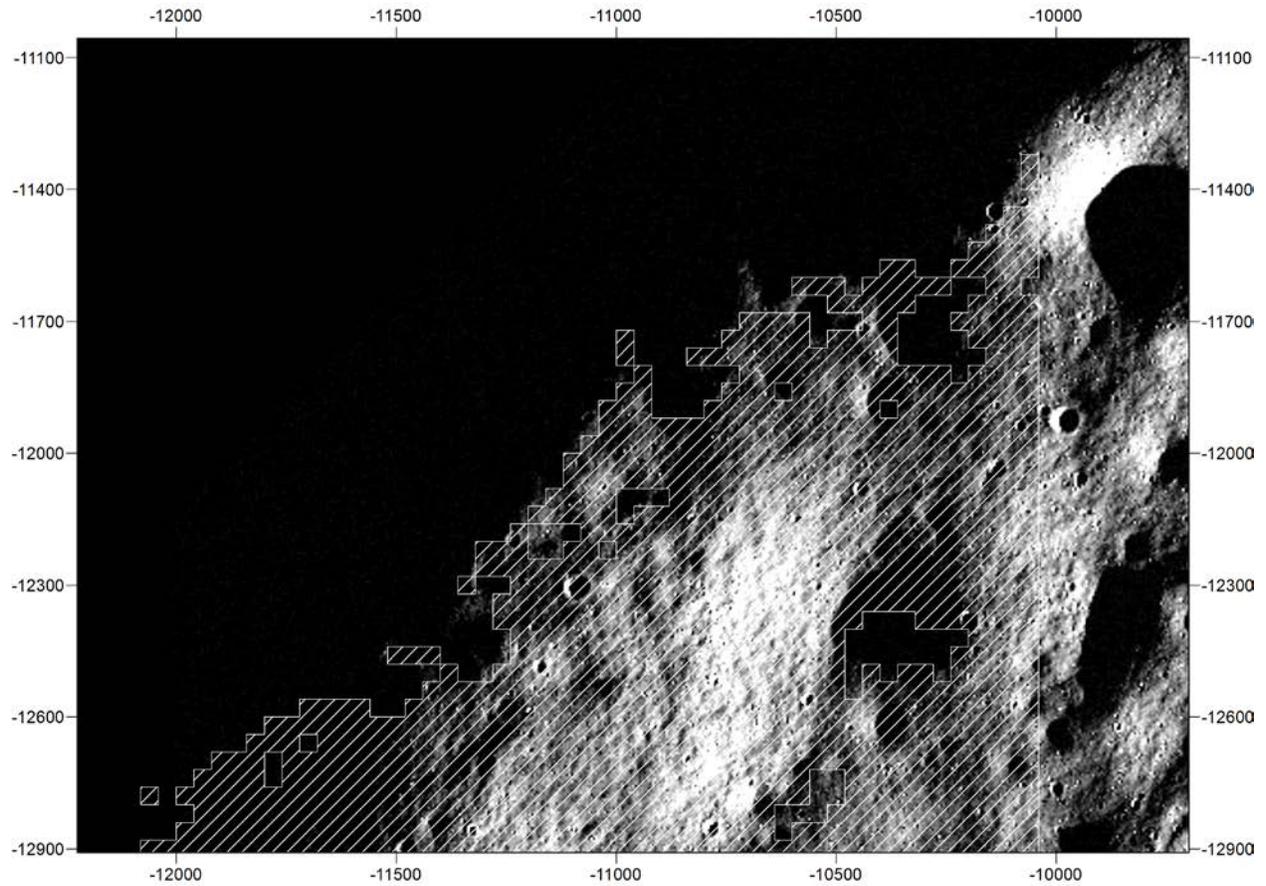

Figure 18. Comparison of the LQCIP maps simulated with the ICAT (a) and Coverage Tool (b), for a sub-area of the Connecting Ridge RoI. The map of the LQCIP difference as simulated by the two tools (c) and an histogram of the differences (d) are also given.

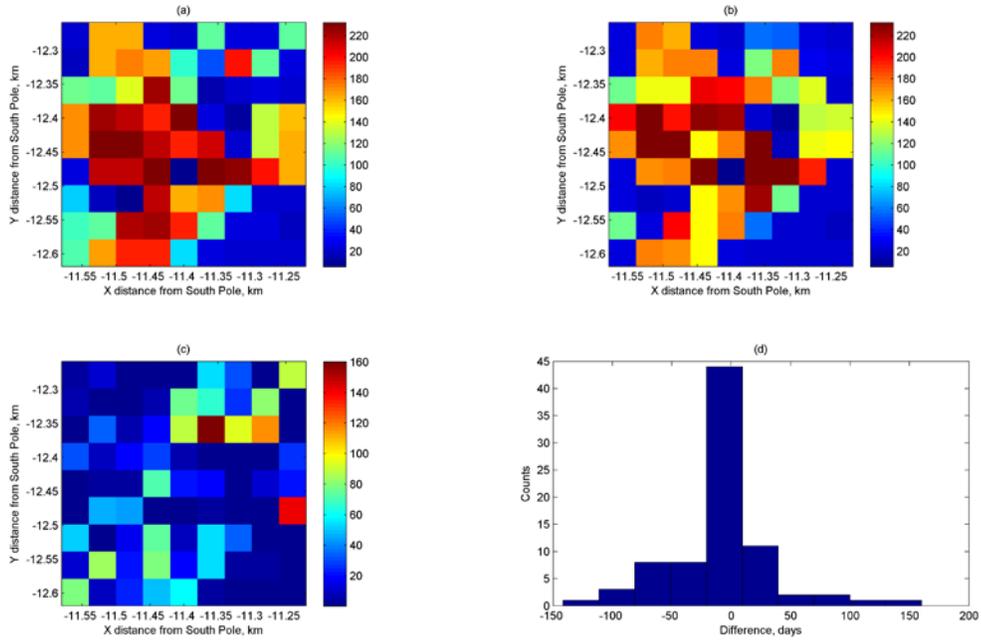

Figure 19. Horizon computed using the ICAT and the Coverage Tool (a), and distance of the corresponding terrain point generating the horizon, computed by the ICAT (b).

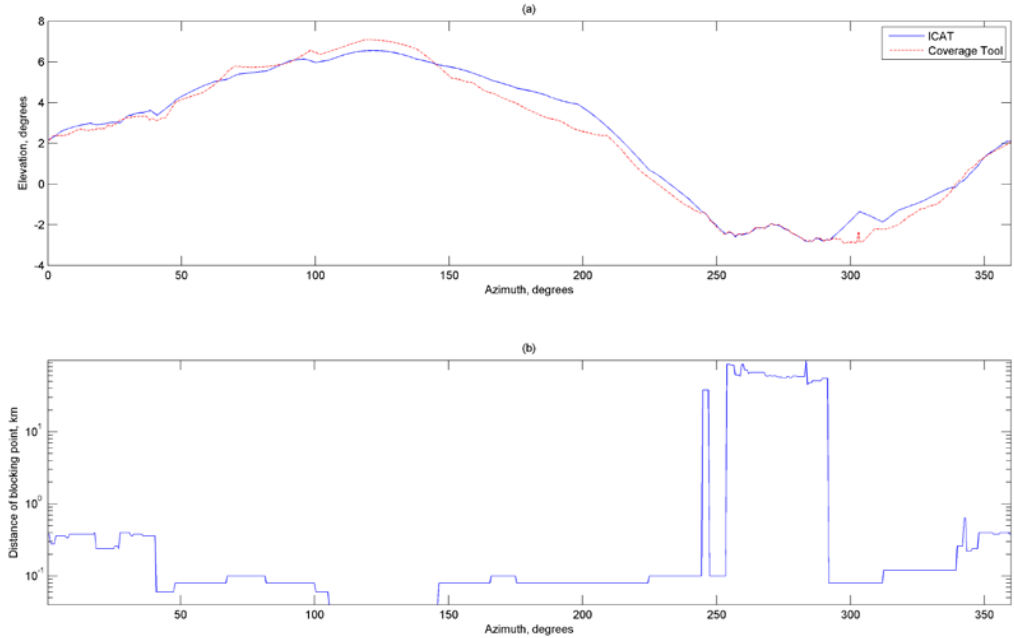

Figure 20. Definition of the landing hazards.

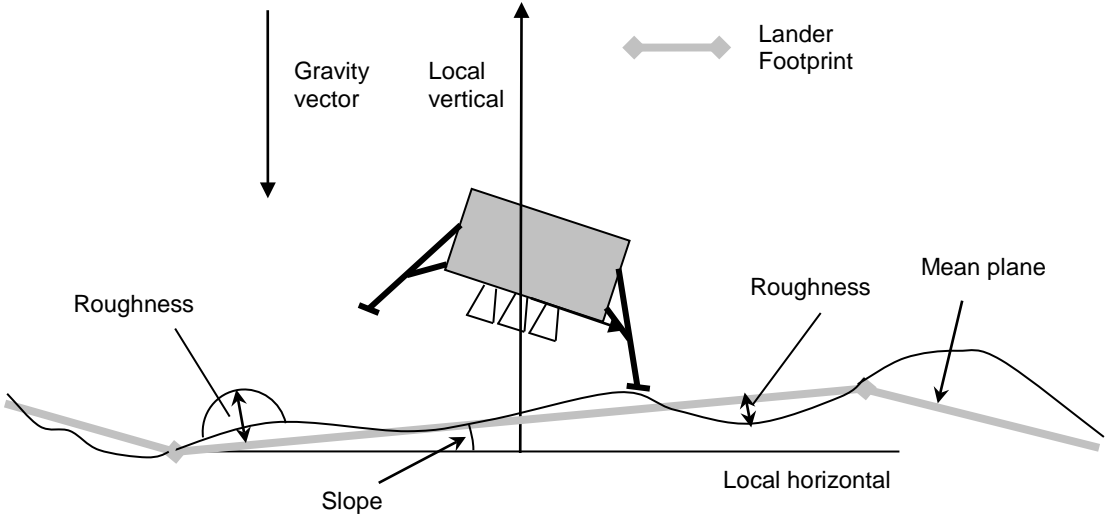

Figure 21. Top: topographic location map for the analysed area in the CR1 RoI (left) and slope maps derived from pre-gridded datasets at various baselength scales (right). Bottom: LOLA EDR point distribution for CR1 (a), Delaunay-triangulation and topography (b), TIN-based slope distribution (c), using the same colour code as in Figure 21. Note that these diagrams indicate an apparent correlation between slope and data coverage, in the sense that sparser coverage results in lower average slopes (gaps may have a size of up to 76 m). This is clearly an artefact introduced by uneven data coverage, and implies that areas sampled with few points that present shallow slopes (<15°) should not necessarily be trusted.

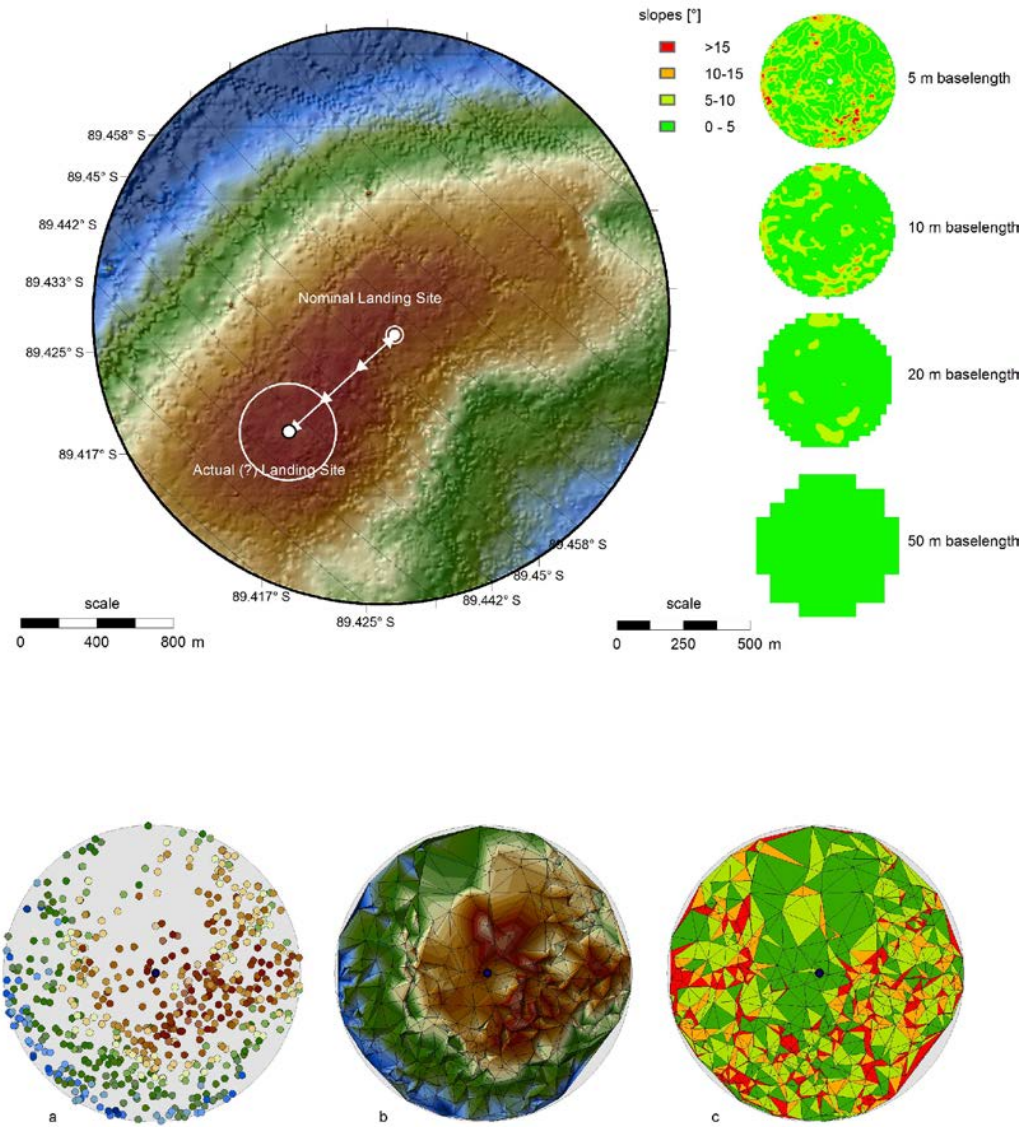

Figure 22. Compilation of slopes based on LOLA GDR analysis for all sites. Mean slopes and standard deviations are given at each grid size. A grid size of only up to 40 m was used for CR1 and CR2 due to the smaller analysis area size (150 m).

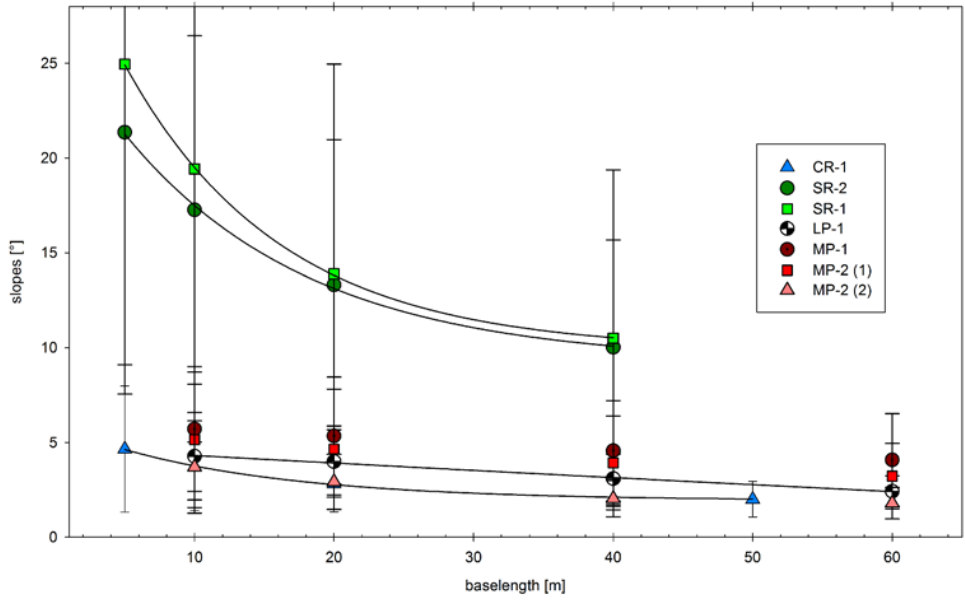

Figure 23. Left: impact-crater measurements for SR1 (2610 craters); grid interval is 200 m. The smallest detected impact crater is 1.2 m in diameter, corresponding to 3 times the pixel scale of the image mosaic used. Right: plots for the corresponding impact-crater size frequency distributions for SR1, for the larger area and for the smaller window analysed. The surface age, estimated from the best-fit curve (Michael and Neukum, 2010), is also given.

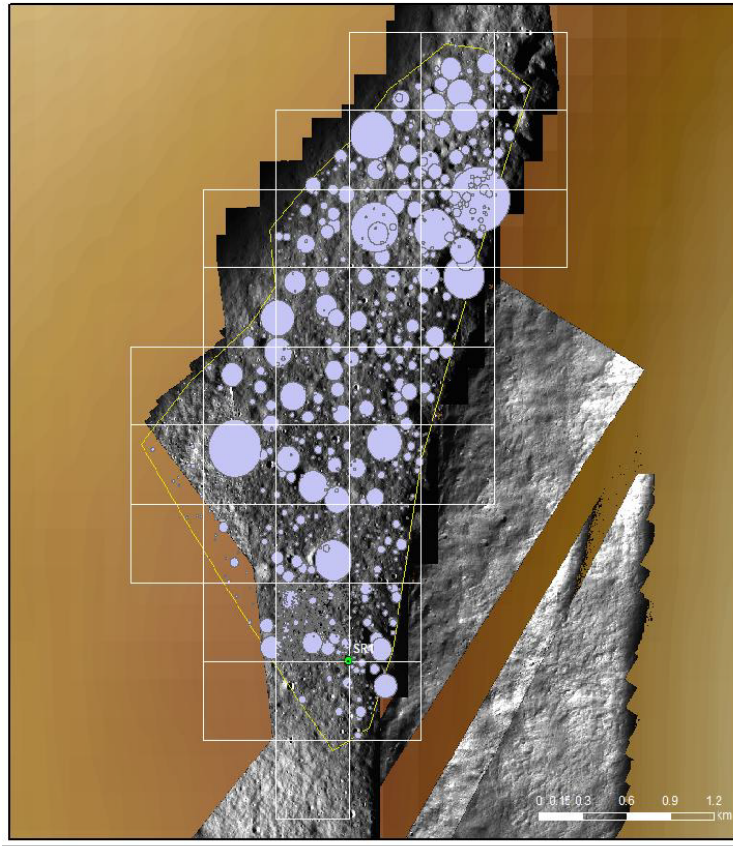
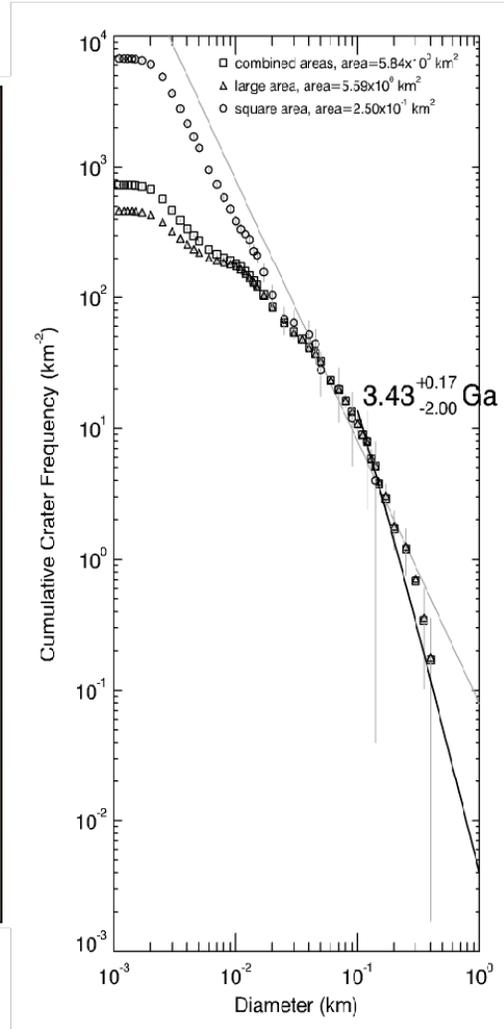

Figure 24. Left: impact-crater measurements for MP1 (2038 craters); grid interval is 200 m and the larger area has a 3 km radius. The smallest detected impact crater is 1.5 m in diameter, corresponding to 3 times the pixel scale of the image mosaic used. Right: plots for the corresponding impact-crater size frequency distributions for MP1, for the larger area and for the smaller window analysed.

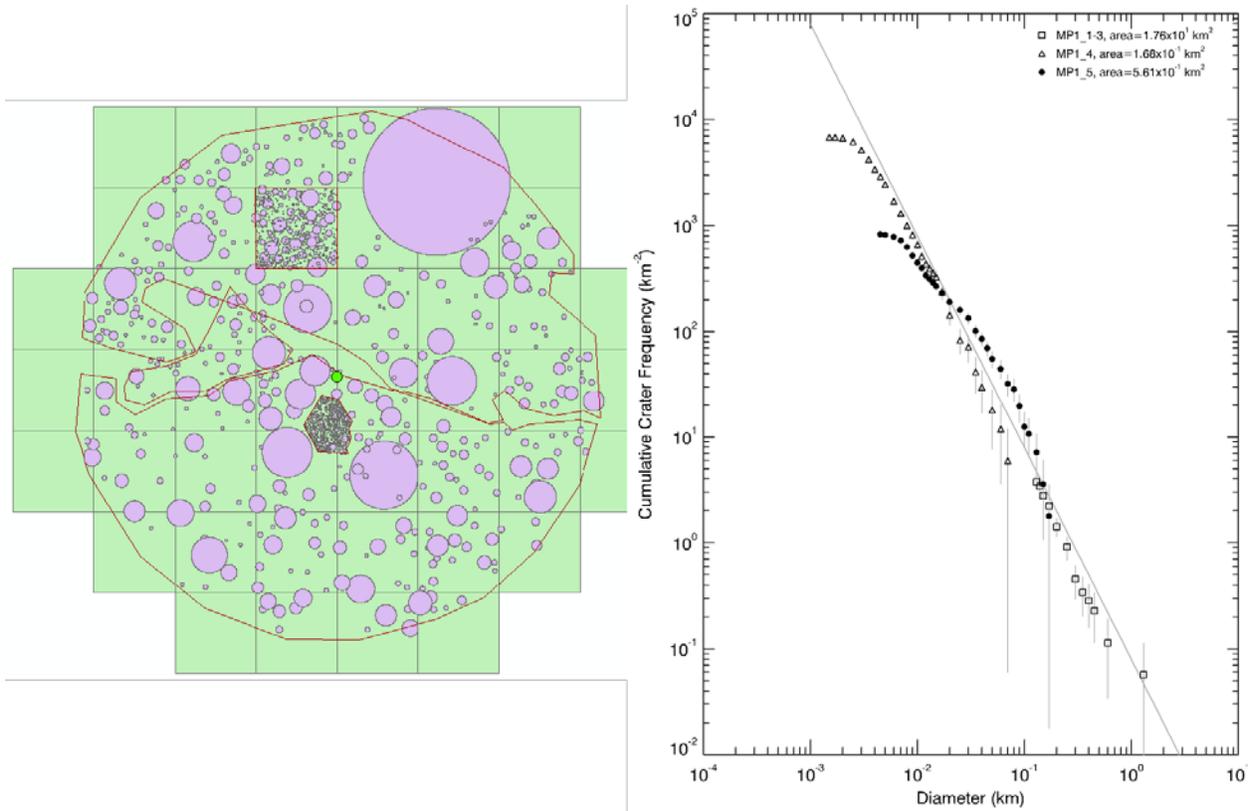

Figure 25. Left: impact-crater measurements for MP2 (2965 craters); grid interval is 200 m and the larger area has a 3 km radius. The smallest detected impact crater is 1.8 m in diameter, corresponding to 3 times the pixel size of the image mosaic used. Right: plots for the corresponding impact-crater size frequency distributions for MP2, for the larger area and for the smaller window analysed. The surface age, estimated from the best-fit curve (Michael and Neukum, 2010), is also given.

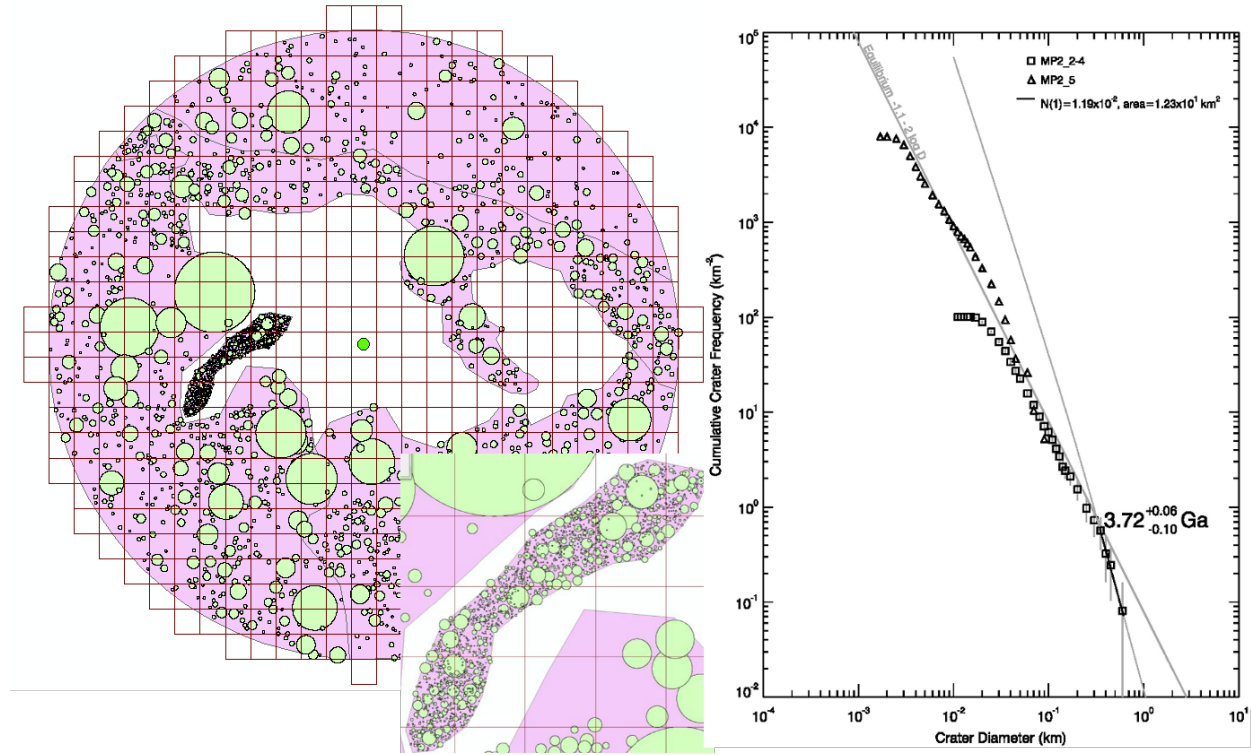

Figure 26. Left: impact-crater measurements for LP1 (3208 craters). The smallest detected impact crater is 2.4 m in diameter, corresponding to 3 times the pixel size of the image mosaic used. Right: plots for the corresponding impact-crater size frequency distributions for SR1, for the larger area (3 km radius) and for the smaller window analysed.

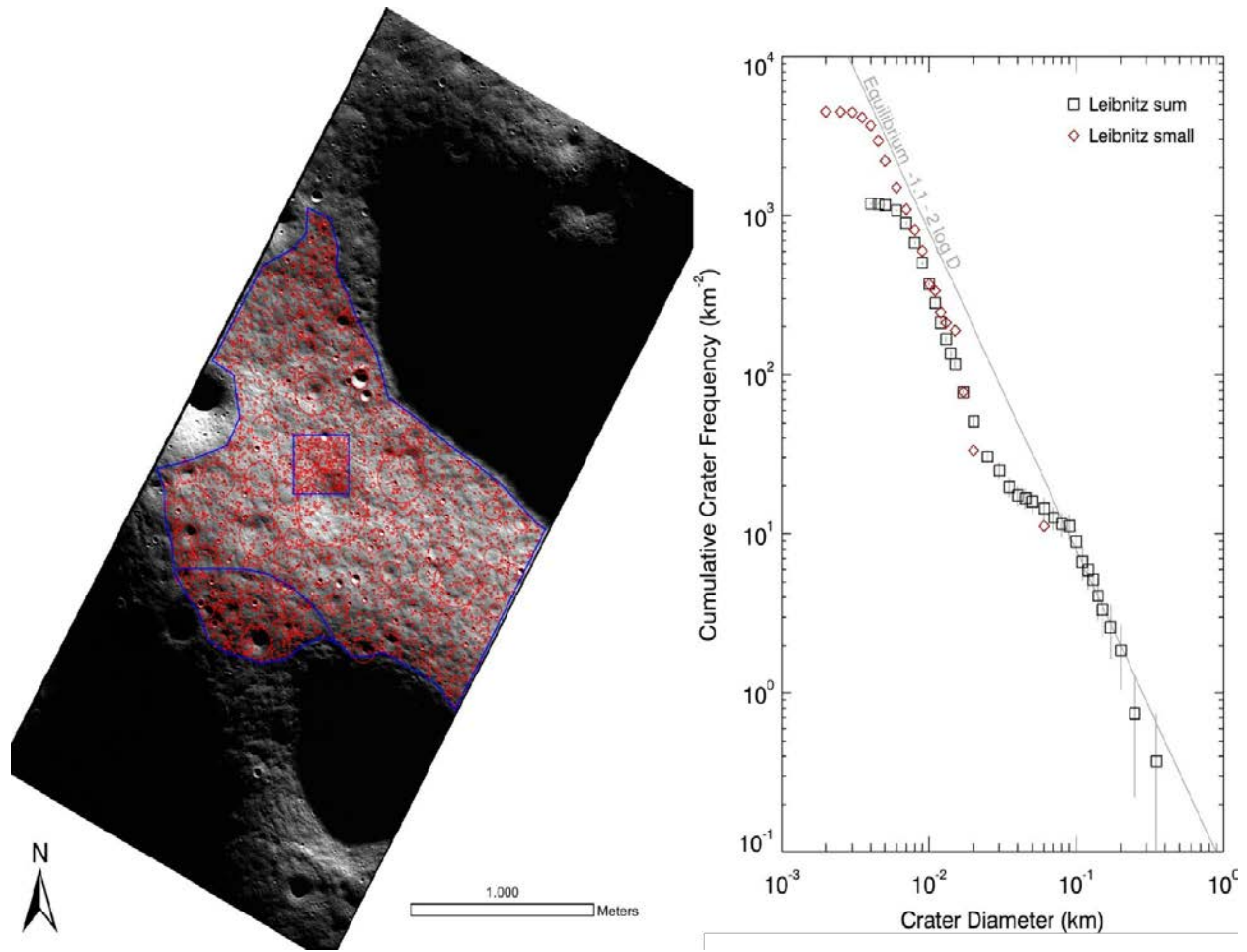

Figure 27. Cumulative logarithmic crater size-frequency distributions for the Apollo 16 landing site as a function of illumination angle. It is clear that at the largest incidence angle (i.e. that which best approaches the polar illumination) a large number of the smallest craters have been missed, although rather larger craters have been identified at this incidence angle.

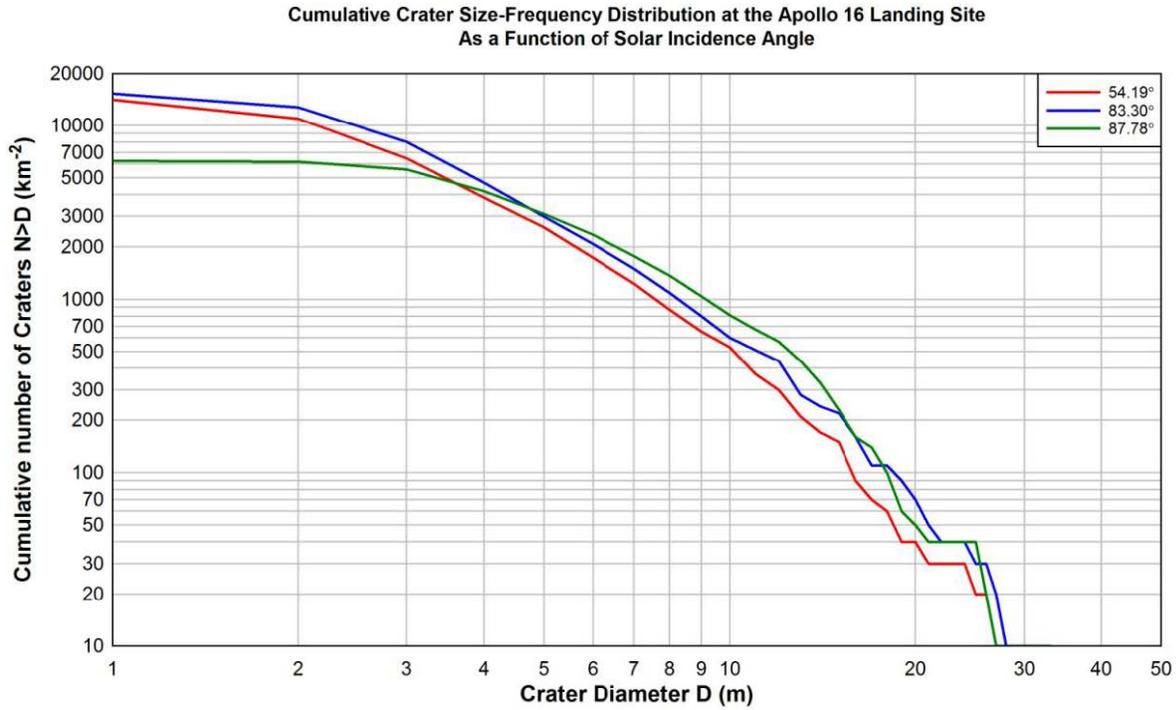

Figure 28. Boulder detection example using LROC NAC image mosaic of the Shackleton rim, around SR1 and SR2 RoI's (top-left), shown with increasing detail (top-right, bottom left and right). Boulders are digitised using a circle representation (bottom-left). The large boulder at the centre of the bottom-right image is 10 m in diameter (for scale).

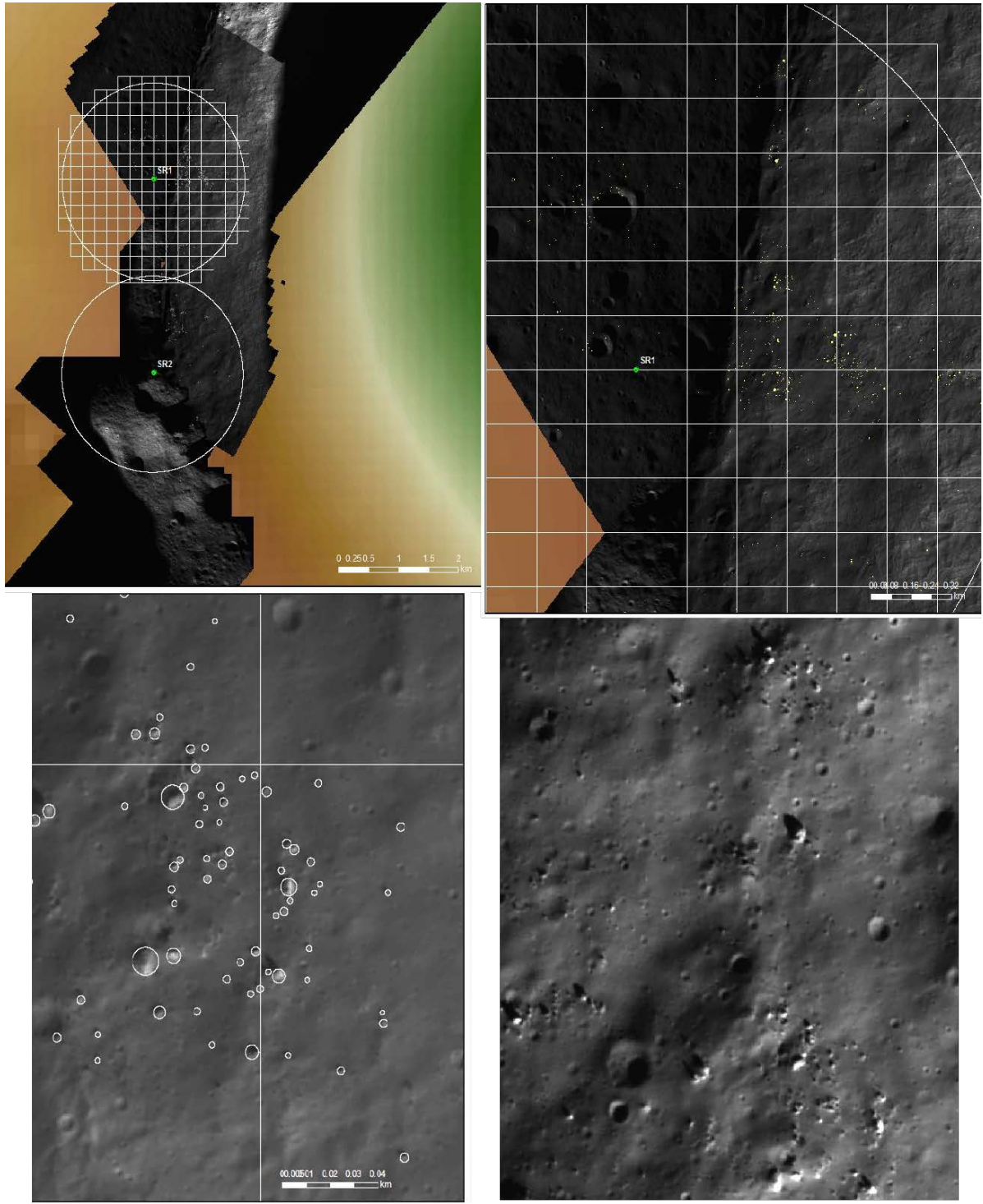

Figure 29. Boulder size-frequency distribution (left) for the 9 km² area around MP2 and for the 4 km by 2.5 km area on the Shackleton crater rim around CR1 and CR2. Boulders map (right) for the SR1/SR2 area.

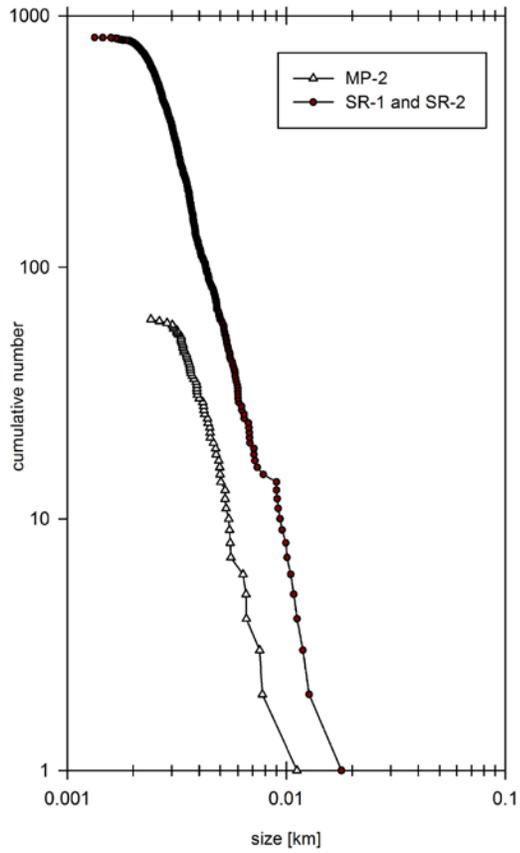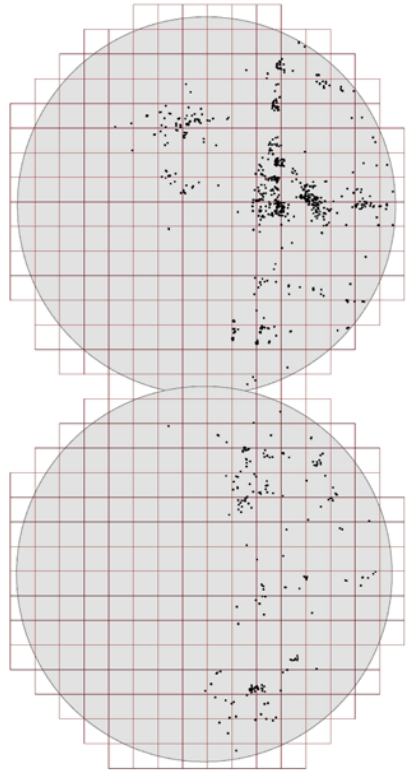

Figure 30. Comparison of the cumulative number per unit area of the craters automatically extracted in the SR1 RoI with the cumulative number per unit area of the craters visually detected in the same RoI (see Sections 3.2)

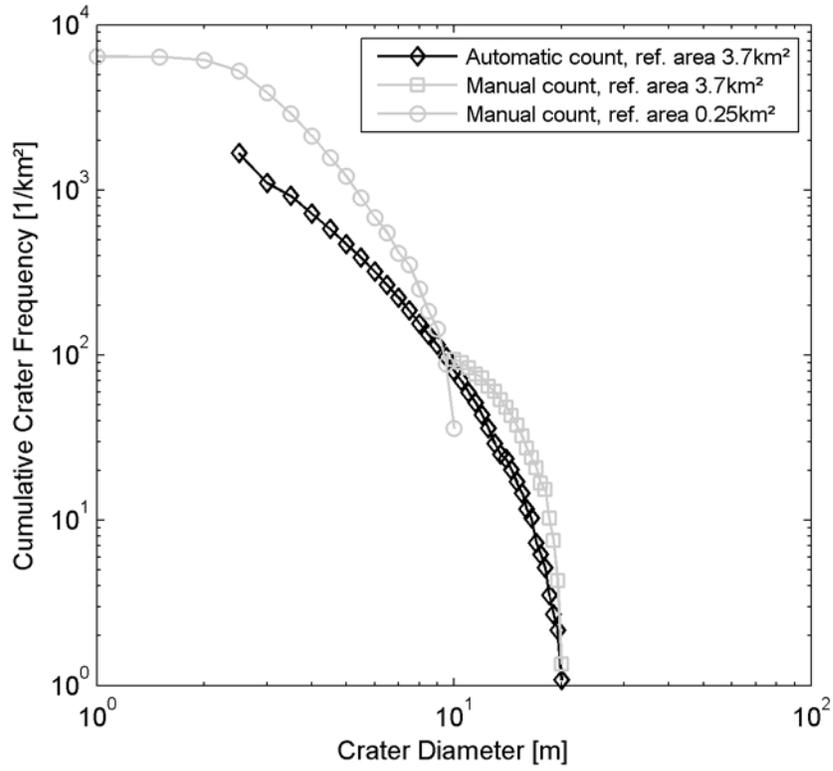

Figure 31. Typical intensity histogram for a NAC image, used for the selection of the threshold used in shadow detection (red line).

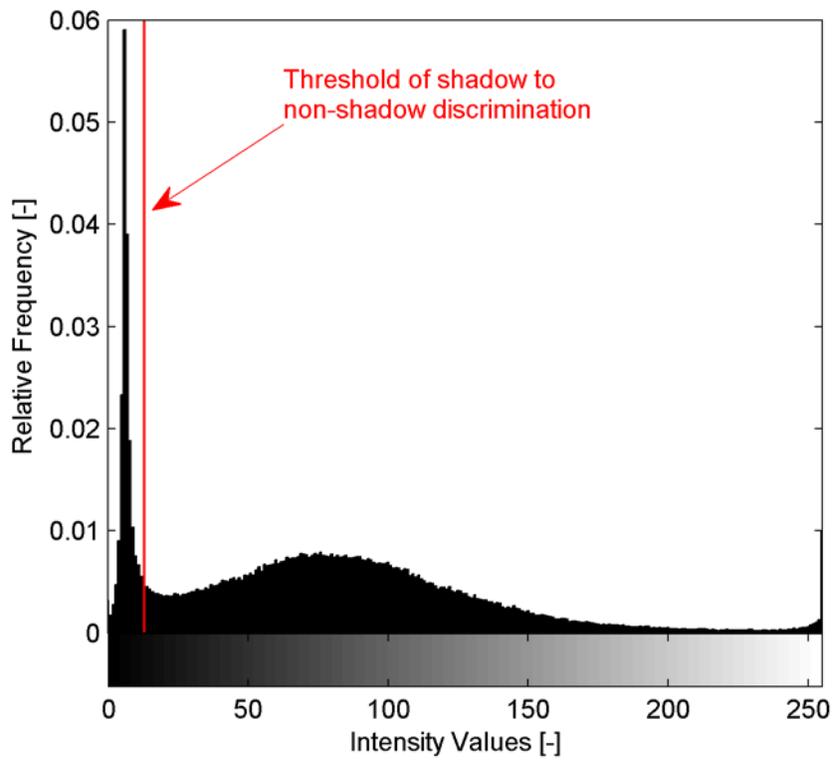

# Tables

Table 1. Lunar southern winter solstice according to method 1 (epoch of maximum sub-solar latitude) and method 2 (epoch halfway through a northbound and southbound equator crossing of the sub-solar point), for the years of interest.

| Method 1 | Method 2 |
|---|---|
| 22/10/2018 | 23/10/2018 |
| 12/10/2019 | 03/10/2019 |
| 30/09/2020 | 15/09/2020 |
| 27/08/2021 | 26/08/2021 |
| 14/08/2022 | 10/08/2022 |

Table 2. Coordinates of the centre point of the Regions of Interest, in the Mean Earth/Polar Axis (ME/PA) reference system (DE421, Williams et al., 2008)

| Region name | ID | Centre of Region of Interest | | | |
|---|---|---|---|---|---|
| | | Geodetic | | Polar stereographic | |
| | | Lat (°) | Lon (°) | X (km) | Y (km) |
| Connecting Ridge | CR1 | -89.4679 | -137.411 | -10.92 | -11.88 |
| de Gerlache Rim | GR1 | -88.6952 | -68.2846 | -36.76 | 14.64 |
| Leibnitz beta Plateau | LP1 | -85.4596 | 31.78053 | 72.55 | 117.1 |
| Malapert Peak | MP1 | -85.9756 | -2.11241 | -4.48 | 121.8 |
| Malapert Peak | MP2 | -86.0265 | 2.757964 | 5.8 | 120.4 |
| Shackleton Rim | SR1 | -89.7755 | -156.448 | -2.72 | -6.24 |
| Shackleton Rim | SR2 | -89.689 | -162.351 | -2.84 | -9 |
| de Gerlache Rim | GR2 | -89.0073 | -94.7636 | -30 | -2.48 |
| Shackleton Rim | SR3 | -89.812 | 52.125 | 4.48 | 3.48 |
| Shackleton Vicinity | SV1 | -88.8247 | 124.1359 | 29.48 | -20 |
| Leibnitz beta Plateau | LP2 | -85.2934 | 37.0304 | 86 | 114 |
| Leibnitz beta Plateau | LP3 | -85.5566 | 37.4649 | 82 | 107 |

Table 3. LOLA data density within the selected Regions of Interest, in percentage of map pixels having at least one measurement at a given LDEM resolution, as a function of coverage from a given latitude to -90° latitude. When a map of a given resolution is not available for a certain RoI, this is indicated with NA.

| Region | Latitude band (from 90°S) | 75° S | 80° S | 87.5° S | 87.5° S |
|:---:|:---:|:---:|:---:|:---:|:---:|
| | Resolution (m) | 60 | 40 | 20 | 10 |
| CR1 | | 99.8 | 97.8 | 72.5 | 27.6 |
| SR1 | | 99.9 | 97.2 | 73.0 | 28.3 |
| GR1 | | 79.3 | 65.4 | 33.0 | 9.6 |
| MP1 | | 56.0 | 41.6 | NA | NA |
| MP2 | | 61.2 | 46.2 | NA | NA |
| SV1 | | 99.4 | 96.1 | 70.3 | 25.8 |
| LP1 | | 63.5 | 46.5 | NA | NA |
| LP2 | | 56.6 | 41.7 | NA | NA |

Table 4. Coordinates of the 5 points in the region -88° to - 90° latitude, receiving the most illumination at the surface, for the period 2010-Mar-31-12:00 to 2010-Sep-24-00:00. The average illumination is also reported.

| Polar stereographic coordinates | | | | Average illumination (%) | |
| --- | --- | --- | --- | --- | --- |
| ESA | | ASP | | ESA | ASP |
| X (km) | Y (km) | X (km) | Y (km) | | |
| -2.64 | -9.12 | -2.64 | -9.12 | 84.1263 | 81.355 |
| 30.24 | -20.16 | 30.24 | -20.16 | 83.16 | 80.5085 |
| -39.60 | 16.08 | 39.60 | 16.08 | 78.64 | 76.2712 |
| -18.96 | -8.64 | 18.96 | -8.64 | 77.34 | 74.5763 |
| -10.80 | -12.00 | -11.52 | -12.24 | 76.25 | 78.8136 |

Table 5. Summary of results of crater counting in the vicinity of the Apollo 16 landing site as a function of solar incidence angle. The areas counted were identical for all three images. Although far fewer craters were observed in the highest incidence angle image (M132732855R) rather more larger craters were counted in this image, causing the total areas covered by recognized craters to be similar in all three images.

| Image | Sun Incidence, ° | No. craters | Crater density, $km^{-2}$ | Fractional area |
|---|---|---|---|---|
| M129187331L | 54.19 | 1435 | 14350 | 23.1% |
| M117392541L | 83.30 | 1546 | 15460 | 28.8% |
| M132732855R | 87.78 | 624 | 6240 | 26.3% |

Table 6. Summary of results of boulder counting in the vicinity of the nominal landing sites.

| Site | Image | Sun Incidence, ° | Area surveyed, km$^2$ | No. boulders | Density, km$^{-2}$ | Fractional area |
|---|---|---|---|---|---|---|
| SR1 | M115517994R | 89.52 | 0.10 | 3 | 30 | 3.8×10-4 |
| SR1/2 | M123118567R | 89.66 | 0.11 | 3 | 27 | 3.4 ×10-4 |
| CR1 | M129050120R | 91.38 | 0.20 | 4 | 20 | 2.5 ×10-4 |
| MP1 | M129328363L | 88.22 | 0.14 | 0 | 0 | 0 |
| MP2 | M113882518R | 86.44 | 0.10 | 2 | 20 | 2.5 ×10-4 |

Table 7. Summary of results of boulder counting in the vicinity of the Apollo 16 landing site as a function of solar incidence angle. The areas counted were identical for all three images (0.1 km$^2$).

| Image | Sun Incidence, ° | 1m | 2m | 3m | 4m | 5m | Total | Density, km$^{-2}$ |
|---|---|---|---|---|---|---|---|---|
| M129187331L | 54.19 | 11 | 1 | 1 | 1 | 0 | 14 | 140 |
| M117392541L | 83.3 | 9 | 3 | 2 | 1 | 0 | 15 | 150 |
| M132732855R | 87.78 | 4 | 5 | 3 | 1 | 1 | 14 | 140 |

Table 8. Quality measures of the automatic identification and classification performances for boulders and craters, derived using as reference data a sub-set of the results of visual detection for the SR1 RoI (see Sections 3.2 and 3.3).

|  | Craters | Boulders |
|---|---|---|
| True Positive | 323 | 7 |
| False Positive | 98 | 60 |
| False Negative | 60 | 2 |
| Detection Percentage (%) | 84 | 77 |
| Branching Factor | 0.3 | 8.6 |